# Atomic-scale 3D structural dynamics and functional degradation of Pt alloy nanocatalysts during the oxygen reduction reaction

Chaehwa Jeong[1†], Juhyeok Lee[1,2,3†], Hyesung Jo[1†], KwangHo Lee[4†], SangJae Lee[4], Colin Ophus[5,6], Peter Ercius[3], EunAe Cho[4*] and Yongsoo Yang[1,7*]

[1] *Department of Physics, Korea Advanced Institute of Science and Technology (KAIST), Daejeon 34141, Republic of Korea*

[2] *Energy Geosciences Division, Lawrence Berkeley National Laboratory, Berkeley, CA 94720, USA*

[3] *National Center for Electron Microscopy, Molecular Foundry, Lawrence Berkeley National Laboratory, Berkeley, CA 94720, USA*

[4] *Department of Materials Science and Engineering, Korea Advanced Institute of Science and Technology (KAIST), Daejeon 34141, Republic of Korea*

[5] *Department of Materials Science and Engineering, Stanford University, Stanford, CA 94305, USA*

[6] *Precourt Institute for Energy, Stanford University, Stanford, 94305, USA*

[7] *Graduate School of Semiconductor Technology, School of Electrical Engineering, Korea Advanced Institute of Science and Technology (KAIST), Daejeon 34141, Republic of Korea*

[†]These authors contributed equally to this work.

*Corresponding author, email: eacho@kaist.ac.kr, yongsoo.yang@kaist.ac.kr

**Pt-based electrocatalysts are the primary choice for fuel cells due to their superior oxygen reduction reaction (ORR) activity. To enhance ORR performance and durability, extensive studies have investigated transition metal alloying, doping, and shape control to optimize the three key governing factors for ORR: geometry, local chemistry, and strain of their surface and subsurface. However, systematic optimization remains incomplete, as it requires an atomic-scale understanding of these factors and their dynamics over potential cycling, as well as their relationship to ORR activity. Here, we implement neural network-assisted atomic electron tomography to measure the 3D atomic structural dynamics and their effects on the functional degradation of PtNi alloy catalysts. Our results reveal that PtNi catalysts undergo shape changes, surface alloying, and strain relaxation during cycling, which can be effectively mitigated by Ga doping. By combining geometry, local chemistry, and strain analysis, we calculated the changes in ORR activity over thousands of cycles and observed that Ga doping leads to higher initial activity and greater stability. These findings offer a pathway to understanding 3D atomic structural dynamics and their relation to ORR activity during cycling, paving the way for the systematic design of durable, high-efficiency nanocatalysts.**

# Main

**Introduction**

Electrocatalysts are crucial substances in electrochemical energy conversion and storage, greatly enhancing the efficiency and rate of reactions in applications such as fuel cells, electrolyzers, and batteries. Especially for polymer electrolyte membrane fuel cells (PEMFCs), the effectiveness of electrocatalysts is essential to improve the slow kinetics of the oxygen reduction reaction (ORR) at the cathode, and its proper development can make it the most attractive power source for zero-emission vehicles due to their environmental benefits, energy efficiency, and power density[1,2].

Traditionally, Pt has been used as a commercial catalyst; however, its high cost and limited availability have driven extensive research into alternative electrocatalyst development through a variety of strategies[3–17]. For example, introducing Ga atoms into the Pt surface induces a downward shift in the d-band through p-d orbital hybridization[14], which enhances the catalytic activity by modulating the binding energies of reaction intermediates such as O and OH. Other approaches including shape control[3–7], Pt-M (M = transition metal, e.g., Ni, Co, Fe) alloys[8–14], dealloying[15,16], and doping[11,17] have also been demonstrated, all aiming to improve ORR activity while maintaining stability under fuel cell operating conditions. These strategies work by tailoring the interplay among geometry[18–20], local chemistry[21,22], and strain[5,23], in both surfaces and subsurface regions. Furthermore, from the perspective of durability, understanding the surface and subsurface structural dynamics during potential cycling and their relation to ORR activity is crucial. In this regard, precise measurement of the three-dimensional (3D) atomic structural dynamics of nanocatalysts is essential to systematically optimize catalyst performance in terms of ORR activity and durability.

Most microscopic and spectroscopic techniques provide either ensemble-averaged information over a large number of nanocatalysts[24–26] or are limited to low-dimensional measurements[17,27–29], rather than offering direct 3D local atomic structural information. Consequently, changes in the 3D atomic arrangement on actual surfaces and subsurfaces of nanocatalysts during potential cycles remain experimentally unexplored. The majority of catalytic studies rely on theoretical models and simulations[9,30–32], which have difficulty accurately describing the physical properties of actual structures, particularly for heterogeneous surfaces and adsorbates[33,34].

Atomic electron tomography (AET) has recently emerged as a powerful technique for 3D structural imaging at the single-atom level[35–38] and has been utilized to analyze Pt-based nanocatalysts[39,40], enabling the identification of active sites and the examination of the effects of surface strain, ligands, and doping[41]. Additionally, neural network-assisted AET has been developed to address artifacts from data imperfections, enhancing the accuracy of atomic structures, particularly on nanocrystal surfaces[39,42].

Here, we implemented neural network-assisted AET to determine the 3D atomic structural dynamics of octahedral PtNi nanocatalysts and clarified the effect of Ga doping in terms of geometry, local chemistry, and strain over potential cycling. By integrating these factors, we calculated how ORR activity changes over thousands of cycles.

## Results

### Determination of 3D atomic structural dynamics of PtNi and Ga-PtNi nanoparticles

PtNi and Ga-doped PtNi (Ga-PtNi) nanoparticles of approximately 8-10 nm in diameter were synthesized in slightly truncated octahedral shape[43–46] via a one-pot synthesis method (Methods). The PtNi and Ga-PtNi particles were drop-cast onto carbon membranes with a thickness of 3-4 nm, followed by potential cycling for various numbers of cycles. The potential cycling was conducted in oxygen ($O_2$)-saturated 0.1 M $HClO_4$ solution, with potentials cycled between 0.6 and 1.0 V vs. reversible hydrogen electrode (RHE) at a scan rate of 100 mV s$^{-1}$. The measurements were carried out using a three-electrode system comprising a glassy carbon electrode, a platinum wire as the counter electrode, and a saturated calomel electrode as the reference electrode (Methods). A total of 8 PtNi nanoparticles [2 pristine and 6 after 12,000 (12k) cycles] and 9 Ga-PtNi nanoparticles (3 pristine, 2 after 4k cycles, 2 after 8k cycles, and 2 after 12k cycles) were selected for AET measurements conducted using annular dark-field scanning transmission electron microscopy (ADF-STEM) mode (see Methods and Supplementary Figs. 1-10). Note that various cycling was performed on different nanoparticles, rather than on a single nanoparticle. After image post-processing and tilt-series alignment, the atomic resolution 3D tomograms of the 8 PtNi and 9 Ga-PtNi nanoparticles were reconstructed using the GENFIRE algorithm[47]. Subsequently, neural network-based volume data augmentation was applied to enhance the reliability of the atomic structures (Methods). The 3D atomic coordinates and chemical species of Pt and Ni atoms in each nanoparticle were identified from the final tomograms using atom tracing and species classification techniques (Methods)[38,40,48]. The average precision of the 3D atomic coordinates across all nanoparticles, as determined from multislice simulations, was calculated to be 31.7 ± 5.2 pm (see Methods and Supplementary Tables 1 and 2). Note that Ga and Ni atoms cannot be distinguished in our ADF-STEM image contrast due to their similarity in atomic number.

### Geometrical dynamics of PtNi and Ga-PtNi nanoparticles

Figure 1a-r depicts the 3D atomic structures and chemical compositions of six representative nanoparticles [one selected from each of the pristine PtNi (Fig. 1a, g, m), 12k cycled PtNi (Fig. 1b, h, n), pristine Ga-PtNi (Fig. 1c, i, o), 4k cycled Ga-PtNi (Fig. 1d, j, p), 8k cycled Ga-PtNi (Fig. 1e, k, q), and 12k cycled Ga-PtNi particles (Fig. 1f, l, r)], clearly illustrating the heterogeneous chemical distribution within the nanoparticles. The pristine nanocatalysts consistently exhibit a Pt-to-Ni ratio of approximately 6:4 for both PtNi and Ga-PtNi (Supplementary Tables 1-3). Previous studies have shown that undoped octahedral PtNi nanocatalysts with this composition undergo substantial Ni leaching, leading to a morphological change from an octahedral to a spherical shape during potential cycles[29,49].

However, among the six PtNi particles measured after 12k cycles, only one particle showed the expected behavior, exhibiting substantial Ni leaching together with notable shape change (Supplementary Tables 1 and 4). We further verified this through EDS measurements of hundreds of PtNi nanoparticles after 12k cycles. As shown in Supplementary Fig. 11, the nanoparticles exhibited distinct differences in compositional changes. While a small fraction (~a few percent) displayed strong evidence of Ni leaching, reaching ~80 at% Pt, the majority retained a composition similar to that of pristine particles (~65 at% Pt). The catalytic activity of nanocrystals during cycling can vary depending on the electrical boundary conditions connecting the catalysts to the electrode, resulting in a considerable fraction of nanoparticles being inactive during the reaction. Therefore, we selected the particle that showed the expected behavior for detailed analysis, and the results from the other particles are also summarized in Supplementary Figs. 12 and 13.

As can be seen in Fig. 1a-l, the PtNi nanocrystals underwent a geometrical change from an octahedral to a more spherical shape (truncated octahedron) during 12k potential cycles. Utilizing the full 3D atomic structure, we quantitatively determined the 3D surfaces of the particles via alpha-shape algorithm and calculated their sphericity (Methods). The result corroborates the qualitative observation of shape change, as described in Fig. 1s. Furthermore, for each surface atom, we calculated the normal vector using the cotangent discretization method[50] to identify the facet it belongs to (Methods). Fig. 1t-v shows how the proportions of {111}, {100}, and {110} facets change over potential cycling. For PtNi, the fraction of {111} facets decreased, while those of the {100} and {110} facets increased, indicating a transformation into a truncated octahedron due to potential cycling. In contrast, for Ga-PtNi, the proportion of the {111} facet remained mostly constant during the cycles, maintaining the octahedral shape.

For PtNi catalysts, octahedral geometry is highly beneficial for catalytic activity, since {111} facets typically show higher ORR activity by a factor of 100 or more compared to other facets. Thus, this observed structural change is considered one of the main causes of degradation[51,52]. Doping with different metallic species such as Mo or Ga has been suggested as a solution[11,17]. Our findings for Ga-PtNi are consistent with previous reports, clearly showing that the octahedral shape is maintained throughout potential cycles up to 12k for all 9 Ga-PtNi particles measured in our study. The {111} facets of Ga-PtNi after 12k cycles exhibit more concave curvature compared to those of other Ga-PtNi particles (Fig. 1r), also consistent with a previous study[29], further supporting the validity of our dynamic measurements.

Additionally, the octahedral shape of nanoparticles can appear spherical in 2D projection-based imaging along certain angles (Supplementary Fig. 14), which can be misleading. This highlights the importance of conducting full 3D structural investigations to truly understand the geometric details of nanocatalysts.

**Local chemistry dynamics of PtNi and Ga-PtNi nanoparticles**

During potential cycling, it is known that transition metals in Pt-based alloy catalysts tend to leach out from the nanocatalyst surface, and doping with metals such as Mo or Ga can mitigate the metal dissolution[11,17]. The averaged elemental composition fraction obtained from our 3D structures shows the expected behavior[17], with a substantial reduction in Ni content from 39% to 22% (a 44% reduction) for undoped PtNi during 12k cycling, while that of Ga-PtNi decreased only slightly from 42% to 35% (a 17% reduction) during the same number of cycles (see Fig. 2a). This enhanced compositional stability of Ga-PtNi was further corroborated by sphericity results and 2D EDS elemental mappings (Fig. 1s, Supplementary Fig. 15, and Supplementary Tables 2 and 5).

One of the main advantages of 3D structural analysis compared to 2D projection-based studies is that the physical properties can be quantitatively profiled as a function of distance from the surface, elucidating the in-depth dynamics of local chemistry. Figure 2b,c shows how the elemental composition changes as a function of distance from the surface during potential cycling for PtNi and Ga-PtNi. In their pristine states, both PtNi and Ga-PtNi exhibit a higher Ni fraction in the subsurface/core regions (> 5 Å from the surface) compared to the surface (< 5 Å from the surface). For undoped PtNi, the Ni fraction decreases in both the surface and subsurface/core regions, with the subsurface/core region losing more Ni compared to the surface region. In the Ga-PtNi case, the depth profile does not show a meaningful change up to 4k cycles. For cycles above 8k, a substantial reduction of Ni is observed only in the subsurface/core region compared to particles that underwent fewer cycles, while the Ni fraction at the surface is maintained. Our results confirm that Ni atoms are leaching out during cycling, and Ga doping can effectively suppress this. The notable difference between PtNi and Ga-PtNi in terms of surface Ni dynamics can be explained by the distribution of Ga dopants. The Ga atoms are expected to be mainly distributed on the surface of the nanocrystals, making it difficult for surface Ni atoms to leach out[17]. Therefore, the surface Ni fraction is mostly preserved even after 12k cycles, while subsurface/core Ni atoms are leached out only through limited surface areas where Ga atoms are absent. This preservation allows the octahedral geometry to be maintained even after extensive cycling. In contrast, the surface Ni atoms of undoped PtNi nanocrystals, lacking the protection of Ga dopants, experience a substantial reduction in Ni fraction at both the surface and subsurface/core levels, leading to geometrical deformation towards a more spherical shape.

Note that the dissolution of subsurface/core Ni atoms over potential cycling has been posited in previous experimental[29] and simulation studies[53]. When Ni dissolution occurs on the surface, subsurface/core atoms move to fill the resulting Ni vacancies. Because Ni atoms are expected to have a lower diffusion activation barrier compared to Pt, subsurface/core Ni atoms will continuously move towards the surface and leach out. Our results directly show this phenomenon at the atomic scale in 3D.

The migration of Ni atoms during cycling not only affects the geometrical deformation of the nanocrystals but also directly influences the ORR activity via the ligand effect[54]. Therefore, based on our 3D atomic structures, we calculated the chemical short-range order parameter (CSRO) for every atom within the nanoparticles to describe their local chemical disorder (Methods). A CSRO value close to 0 indicates that the local region around a given atom exhibits the chemical composition of the overall nanoparticle. Values greater than 0 suggest local chemical segregation, and values less than 0 indicate a degree of alloying higher than the global composition. Undoped PtNi exhibits Pt CSRO values of about 0.3 at the surface, indicating a strong tendency for Pt segregation (Fig. 2d). Moving towards the interior of the particle, the Pt CSRO approaches 0, suggesting that the local composition becomes similar to the global composition. After 12k cycles, the Pt CSRO at the surface decreases to less than half of the value of the pristine particle (indicating alloying), while the Pt CSRO of the subsurface/core region becomes substantially higher than before cycling (indicating segregation).

Pristine Ga-PtNi exhibits behavior similar to undoped pristine PtNi, with surface Pt segregation and a more alloyed subsurface/core (Fig. 2e). As the number of cycles increases, the surface Pt CSRO gradually decreases (indicating more alloyed configurations), while the subsurface/core Pt CSRO gradually increases (indicating more segregated configurations). The difference in the Pt CSRO depth profile between pristine and 12k cycled Ga-PtNi particles is less pronounced than in undoped PtNi. After 12k cycles, the surface and subsurface/core Pt CSRO values of Ga-PtNi are at the same level, whereas undoped PtNi shows an inverted behavior in Pt CSRO for the surface and subsurface/core regions between the pristine and 12k cycled particles.

We expect that the CSRO of Ni will behave approximately opposite to the Pt CSRO, as shown in Fig. 2f,g. In the pristine state, Ni exists in a more alloy-like form compared to Pt at the surface. However, subsurface/core Ni tends to be in a more segregated state. As cycling progresses, this trend changes, with subsurface/core Ni showing an alloy composition consistent with the entire nanoparticle. This indicates that segregated Ni atoms are diffusing towards the surface during cycling. Note that Ni CSRO and Pt CSRO are not completely complementary; the CSRO for Pt, which has a higher global composition, is more sensitive to local chemistry, as shown in Fig. 2d-g.

**Strain dynamics of PtNi and Ga-PtNi nanoparticles**

Strain is another crucial factor in determining the performance of a catalyst. Strain alters the electronic structure, resulting in changes in adsorption energies. Pt nanoparticles are expected to show their optimal ORR activity at approximately 3% compressive surface strain[17,55]. We calculated the volumetric strains of Pt atoms by comparing the measured local lattice constant with that of bulk Pt (Methods)[56]. As shown in Fig. 3a, both PtNi and Ga-PtNi exhibit compressive surface strain in their pristine state, with Ga-doped particles showing stronger compressive strain closer to the target value of 3%, indicating that Ga doping is beneficial for improving the catalytic performance.

As the cycling continues, the strain undergoes substantial relaxation. As shown in Fig. 3a, both PtNi and Ga-PtNi nanoparticles exhibit reductions in average compressive surface strain between the pristine and 12k-cycled states. For PtNi, the surface strain decreases by approximately 40.9% (from −2.2% to −1.3%), while for Ga-PtNi, it decreases by 26.1% (from −2.3% to −1.7%). Although the difference between the two 12k-cycled cases lies near the boundary of the error bars, this result is consistent with the findings from the geometrical and chemical analyses: Since Pt has a larger lattice constant than Ni, the dissolution of surface and subsurface/core Ni over the cycles results in strain relaxation toward a less compressive configuration[56]. Again, since Ga doping effectively prevents the dissolution of Ni atoms, the strain relaxation for Ga-PtNi is also suppressed compared to the undoped particles.

We also quantitatively profiled the strain as a function of distance from the surface, as illustrated in Fig. 3b,c. For undoped PtNi, strain relaxation occurs in both surface and subsurface/core regions during potential cycles, indicating that the Ni dissolution occurs throughout the nanoparticle (Fig. 3b). However, for Ga-PtNi, strain relaxation is observed only in the subsurface/core region with minimal changes in the surface (Fig. 3c). This suppressed strain relaxation indicates that Ga doping mitigates Ni dissolution from the surface, aligning with the observed local chemistry dynamics (Fig. 2).

## Calculation of the ORR activity considering geometry, local chemistry, and strain altogether

Finally, we relate the geometry, local chemistry, and strain to calculate the local ORR activity $[\ln(j/j_{Pt(111)})]$ of all surface Pt atoms (Methods). By simultaneously considering geometry, local chemistry, and strain through generalized coordination number (GCN)[57], we derive a modified alloy-sensitive GCN for each surface Pt atom as a descriptor for calculating ORR activity from our 3D atomic structures (see Fig. 4a-f and Methods). As shown in Fig. 4a-f, a clear inhomogeneity can be observed from the ORR activity distribution. Moreover, the {111} facets exhibit better ORR activity compared to those of the {100} or {110} facets, consistent with previous studies[8].

By examining the average values of each nanoparticle's $j/j_{Pt(111)}$ over cycles, we clarified the effect of potential cycling on the overall ORR behavior for PtNi and Ga-PtNi (Fig. 4g). As the cycles progress, the ORR activity decreases for both PtNi and Ga-PtNi. Specifically, the average value of $j/j_{Pt(111)}$ for PtNi shows a decrease of about 17%, which can be attributed to a combinatory effect arising from the decreased fraction of {111} facets, loss of surface Ni atoms, and strain relaxation. In contrast, the ORR activity of Ga-PtNi remains considerably more stable during potential cycling, showing only about 4% reduction even after 12k cycles. Notably, for Ga-PtNi, a slight increase in ORR activity is observed after 4k and 8k cycles compared to the pristine state, albeit within the error margin. While no pronounced strain changes are visible, both an increase in GCN (indicative of a more concave surface curvature) and surface Pt enrichment are evident at these cycles (Supplementary Fig. 16), both of which are known to enhance ORR activity[29,58,59,60]. As discussed earlier, Ga doping effectively modulates the interplay of geometry, chemistry, and strain, helping to maintain a more favorable configuration during cycling. Our descriptor accounts for these factors, successfully predicting the expected ORR behavior.

The predicted ORR activity and durability of PtNi and Ga-PtNi nanoparticle catalysts were experimentally evaluated (Methods). Due to the limited catalyst-to-Nafion ionomer ratio, oxygen transport through the Nafion ionomer can be hindered in the absence of a carbon support. To enable a detectable ORR current, we incorporated carbon support into the system. ORR polarization curves (Supplementary Fig. 17a,b) and cyclic voltammograms (Supplementary Fig. 17c,d) were recorded using the carbon-supported nanoparticles (PtNi/C and Ga-PtNi/C) before and after 12k potential cycles. Specific activities derived from these measurements for pristine particles (Supplementary Fig. 18 and Methods) revealed that Ga-PtNi/C exhibited a superior specific activity of 9.33 mA cm$^{-2}$, outperforming PtNi/C, which showed 3.43 mA cm$^{-2}$. After 12k potential cycles, undoped PtNi/C suffered a 48% decrease in ORR activity (down to 1.78 mA cm$^{-2}$), whereas Ga-PtNi/C exhibited greater durability, retaining 69% of its initial activity with a final value of 6.42 mA cm$^{-2}$. These findings clearly demonstrate that Ga doping enhances both ORR activity and durability of octahedral PtNi nanoparticles, in agreement with atomic-scale structural insights obtained via AET.

**Discussion**

In this study, we observed the changes in ORR activity of PtNi and Ga-PtNi nanoparticles throughout potential cycling by combining three crucial factors for catalytic activity: geometry, local chemistry, and strain. To achieve this, we utilized neural network-assisted AET to determine the full 3D atomic structures of several nanoparticles at various stages of cycling.

Geometric analysis showed that PtNi catalysts experience unfavorable shape changes from octahedral to spherical due to cycling, which can be mitigated by Ga doping. Analysis of the local chemical composition indicated that PtNi catalysts undergo surface alloying which negatively impact ORR activity, and Ga doping prevents this effect. In terms of strain, PtNi catalysts shift towards less compressive surface strain, which is detrimental to ORR activity, whereas Ga doping helps maintain the favorable compressive strain throughout cycling. By integrating these three factors—geometry, local chemistry, and strain—we determined the ORR activity of each nanoparticle at the single-atom level, enabling us to identify the average catalytic activity of each nanoparticle and observe the distinct degradation of this activity over multiple cycles without Ga doping.

Our approach will enable the efficient design of high-performance nanocatalysts through a fundamental understanding of the relationship between surface structure and properties at the atomic scale.

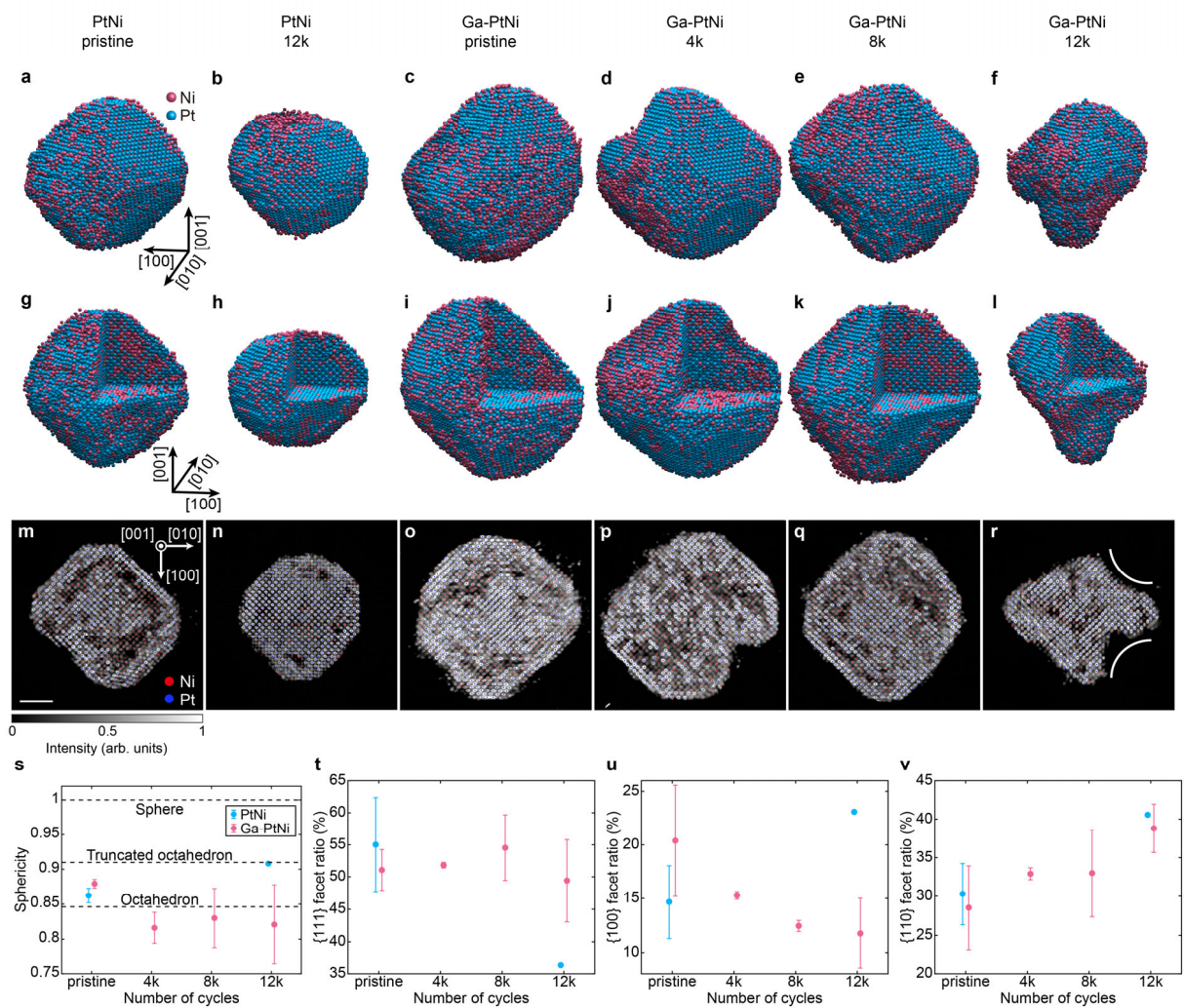

**Figure 1 | Experimentally determined 3D atomic structures of representative PtNi and Ga-PtNi nanoparticles during potential cycles and geometric characterization. a-f**, Overall 3D atomic structure of different PtNi [pristine (**a**) and 12k (**b**)] and Ga-PtNi [pristine (**c**), 4k (**d**), 8k (**e**), and 12k (**f**)] nanoparticles after the indicated number of potential cycles. **g-l**, 3D atomic structures after a 180-degree rotation along the [001] axis from those in (**a-f**), with one octant of the nanoparticles removed to reveal the internal atomic structure. Each image in (**g-l**) corresponds to the images in the first row (**a-f**). **m-r**, 1 Å thick internal slices at the center of the 3D tomograms perpendicular to the [001] direction. The intensity is shown in grayscale (linearly scaled and offset-adjusted to optimize contrast between Pt and Ni atoms), with the atomic coordinates of Pt and Ni marked by blue and red dots, respectively. Concave {111} facets (marked with white lines) are observed in the Ga-PtNi nanoparticle after 12k cycles (**r**). Scale bar, 2 nm. Each image in (**m-r**) corresponds to the images in the first row (**a-f**). Note that, the representative nanoparticles shown in (**a-r**) are PtNi-pristine-p2 (**a, g, m**), PtNi-12k-p1 (**b, h, n**), Ga-PtNi-pristine-p1 (**c, i, o**), Ga-PtNi-4k-p1 (**d, j, p**), Ga-PtNi-8k-p2 (**e, k, q**), and Ga-PtNi-12k-p2 (**f, l, r**) (Supplementary Tables 1 and 2). **s**, Sphericity of the PtNi and Ga-PtNi nanoparticles during the potential cycles. The sphericity values of the ideal sphere, truncated octahedron, and octahedron are marked by dotted lines. **t-v**, Fraction of {111} (**t**), {100} (**u**), and {110} (**v**) facets during potential cycles. Note that the values and error bars for each cycle corresponding to (**s-v**) represent the averages and standard deviations obtained from 2 PtNi pristine, 3 Ga-PtNi pristine, 2 Ga-PtNi 4k, 2 Ga-PtNi 8k, and 2 Ga-PtNi 12k samples. We have only one particle for PtNi 12k, and the error bar cannot be defined in the same manner in this case. Source data for dot or line plots in this figure are provided as a Source Data file.

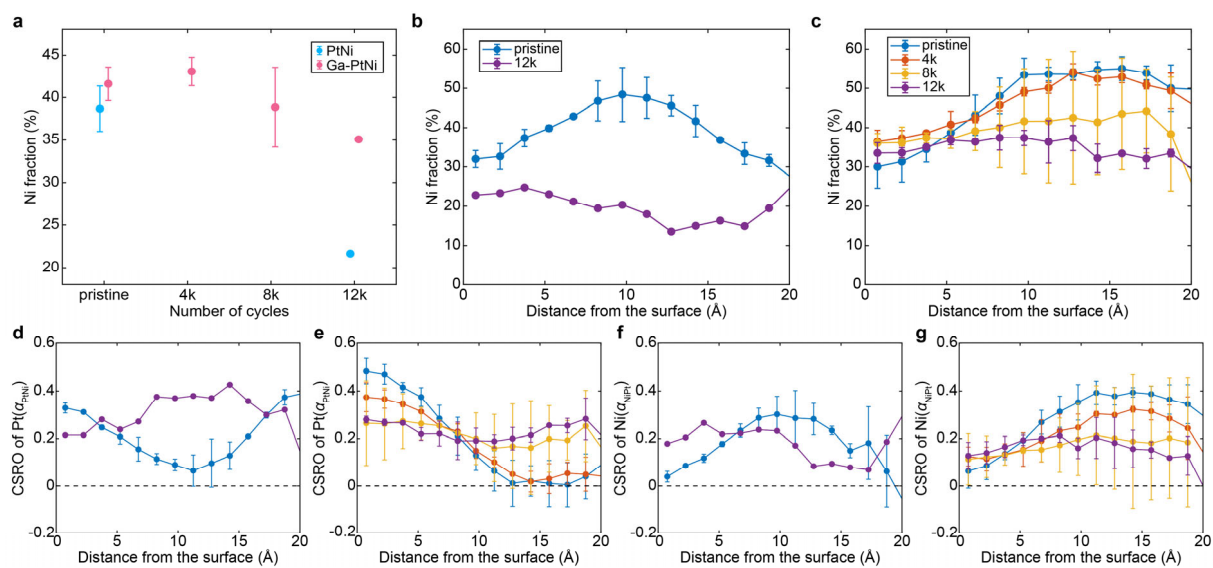

**Figure 2 | Quantitative characterization of the chemical compositions of PtNi and Ga-PtNi nanoparticles during potential cycles. a**, Ni fraction of PtNi and Ga-PtNi nanoparticles during potential cycles. **b**, **c**, Characterization of the Ni fraction as a function of distance from the surface during potential cycles for PtNi (**b**) and Ga-PtNi (**c**) nanoparticles. **d**, **e**, Chemical short-range order parameter (CSRO) of Pt ($\alpha_{PtNi}$) as a function of distance from the surface during potential cycles for PtNi (**d**) and Ga-PtNi (**e**) nanoparticles. **f**, **g**, CSRO of Ni ($\alpha_{NiPt}$) as a function of distance from the surface during potential cycles for PtNi (**f**) and Ga-PtNi (**g**) nanoparticles. A CSRO value close to 0 indicates that the local region around the atom reflects the overall nanoparticle's chemical composition (represented by a dotted line), while values greater than 0 suggest local chemical segregation and values less than 0 indicate a higher degree of alloying than the global composition. Note that the values and error bars for each cycle corresponding to (**a-g**) represent the averages and standard deviations obtained from 2 PtNi pristine, 3 Ga-PtNi pristine, 2 Ga-PtNi 4k, 2 Ga-PtNi 8k, and 2 Ga-PtNi 12k samples. We have only one particle for PtNi 12k, and the error bar cannot be defined in the same manner in this case. Source data for dot or line plots in this figure are provided as a Source Data file.

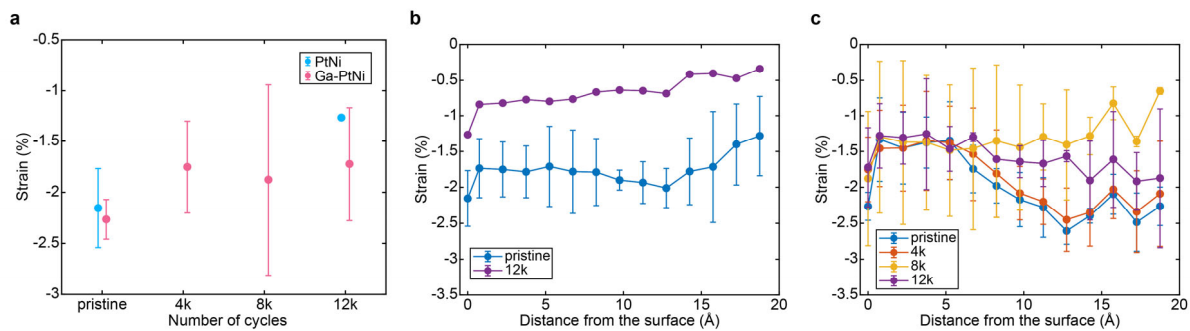

**Figure 3 | Quantitative characterization of the strain of PtNi and Ga-PtNi nanoparticles during potential cycles. a**, Local volumetric strain of surface Pt atoms during potential cycles of PtNi and Ga-PtNi nanoparticles. **b**, **c**, Strain of Pt as a function of distance from the surface during potential cycles for PtNi (**b**) and Ga-PtNi (**c**) nanoparticles. Note that the values and error bars for each cycle corresponding to (**a-c**) represent the averages and standard deviations obtained from 2 PtNi pristine, 3 Ga-PtNi pristine, 2 Ga-PtNi 4k, 2 Ga-PtNi 8k, and 2 Ga-PtNi 12k samples. We have only one particle for PtNi 12k, and the error bar cannot be defined in the same manner in this case. Source data for dot or line plots in this figure are provided as a Source Data file.

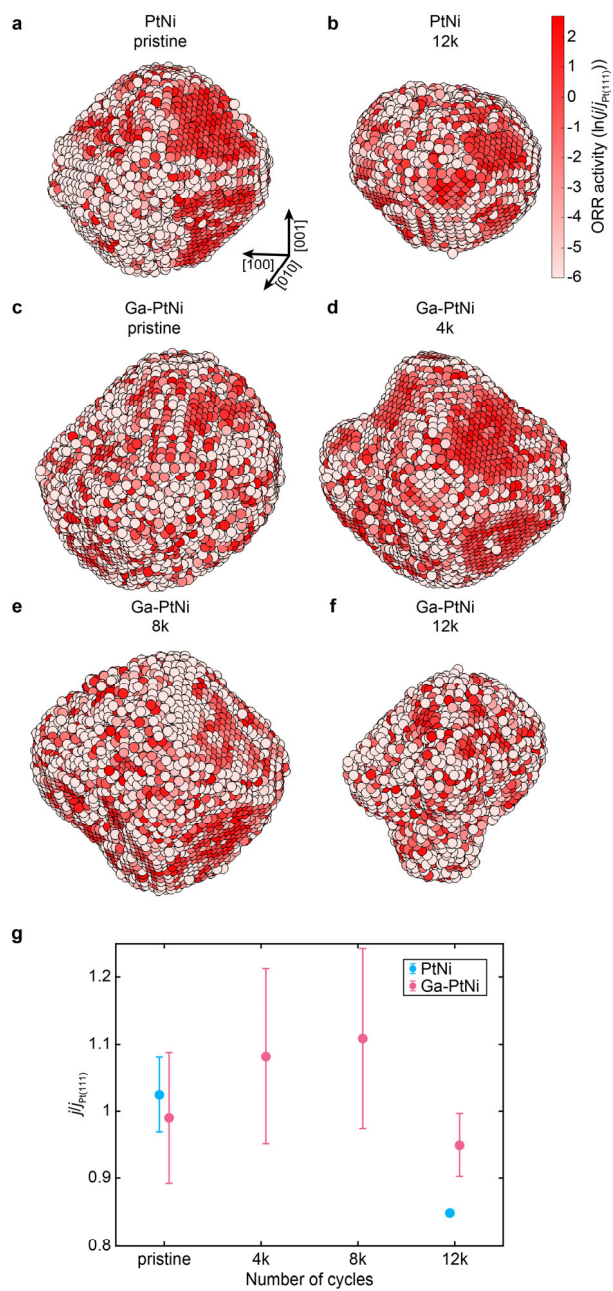

**Figure 4 | Quantitative characterization of ORR activity of PtNi and Ga-PtNi nanoparticles during potential cycles. a-f**, The 3D ORR activity $[\ln(j/j_{Pt(111)})]$ distribution of pristine PtNi (**a**), 12k cycled PtNi (**b**), pristine Ga-PtNi (**c**), 4k cycled Ga-PtNi (**d**), 8k cycled Ga-PtNi (**e**), and 12k cycled Ga-PtNi (**f**). Note that, the representative nanoparticles shown in (**a-f**) are PtNi-pristine-p2, PtNi-12k-p1, Ga-PtNi-pristine-p1, Ga-PtNi-4k-p1, Ga-PtNi-8k-p2, and Ga-PtNi-12k-p2 (Supplementary Tables 1 and 2). **g**, Average of $(j/j_{Pt(111)})$ during potential cycles of PtNi and Ga-PtNi nanoparticles. Note that the values and error bars for each cycle corresponding to (**g**) represent the averages and standard deviations obtained from 2 PtNi pristine, 3 Ga-PtNi pristine, 2 Ga-PtNi 4k, 2 Ga-PtNi 8k, and 2 Ga-PtNi 12k samples. We have only one particle for PtNi 12k, and the error bar cannot be defined in the same manner in this case. Source data for dot or line plots in this figure are provided as a Source Data file.

## Methods

### Sample preparation

*Chemicals and materials*

Platinum(II) acetylacetonate (Pt($C_5H_7O_2$)$_2$, 97%), nickel(II) acetylacetonate (Ni($C_5H_7O_2$)$_2$, 95%), gallium(III) acetylacetonate (Ga($C_5H_7O_2$)$_3$, 99.99%), perchloric acid ($HClO_4$, 70%), and Nafion ionomer solution (20wt%) were purchased from Sigma Aldrich. N,N-dimethylformamide ($C_3H_7NO$, 99.5%) was obtained from Junsei Chemical. Carbon black (Vulcan carbon, XC-72R) was purchased from Cabot Corporation. Acetone ($C_3H_6O$, 99.0%), ethanol ($C_2H_6O$, 99.0%), and 2-propanol ($C_3H_8O$, 99.9%) were purchased from Samchun Chemicals. Deionized water (18.2 MΩ cm) was prepared using a Millipore Direct-A 5 UV. A glassy carbon electrode (GCE), a saturated calomel electrode (SCE), and a platinum mesh were purchased from Fuel Cell Store.

*Synthesis of PtNi*

32 mg of platinum(II) acetylacetonate and 16 mg of nickel(II) acetylacetonate were dissolved in N,N-dimethylformamide with vigorous stirring for 1 hour. The mixture was then heated to 130 °C, maintained at that temperature for 42 hours, and naturally cooled to room temperature. After natural cooling, the precipitate was centrifuged with a mixture of ethanol and acetone and then dried at 60 °C in a vacuum overnight. The synthesized PtNi nanoparticles exhibit a random alloy structure, which is characteristic of the low-temperature solvothermal synthesis process[17,61,62].

*Synthesis of Ga-PtNi*

The synthesis procedure was the same as that of PtNi noted above. Subsequently, 6 mg of gallium(III) acetylacetonate was added to the mixture, which was reheated to 130 °C and maintained for another 42 hours. After natural cooling, the precipitate was centrifuged with a mixture of ethanol and acetone and then dried at 60 °C in a vacuum overnight. These synthesis conditions are known to result in approximately 1.4% Ga doping[17]. The Ga distribution and concentration obtained from EDS analysis are provided in Supplementary Fig. 19 and Supplementary Table 6. The EDS analysis shows an average Ga concentration of approximately 1.28 at% (Supplementary Table 6).

*Synthesis of PtNi/C and Ga-PtNi/C*

To evaluate the ORR performance, carbon-supported PtNi and Ga-PtNi catalysts (PtNi/C and Ga-PtNi/C, respectively) were additionally prepared. Specifically, 80 mg of carbon black was dispersed into the precursor mixtures used for the nanocatalysts, resulting in a Pt composition of 10 wt.%. All other synthesis steps were carried out identically to those used for the preparation of PtNi and Ga-PtNi.

### Characterizations of PtNi/C and Ga-PtNi/C catalysts

The crystal structure of PtNi/C and Ga-PtNi/C catalysts was analyzed using a high-resolution powder X-ray diffractometer (XRD, SmartLab, RIGAKU) with Cu K$_{\alpha 1}$ emission source at the KAIST Analysis Center for Research Advancement (KARA). The XRD patterns of the catalysts (Supplementary Fig. 20) were analyzed to characterize the carbon-supported structures and alloy Pt lattices. Both PtNi/C and Ga-PtNi/C exhibited positively shifted Pt peaks only (Supplementary Fig. 20), confirming alloy formation rather than the presence of separate Ni or Ga phases. The catalyst morphology was examined using a 200 kV field emission TEM at KARA (Talos F200X, Thermo Fisher Scientific) in bright-field imaging mode at two different magnifications (Supplementary Fig. 21). The TEM images revealed well-defined octahedral nanoparticles uniformly dispersed on carbon black.

**Electrochemical characterization and potential cycling of PtNi and Ga-PtNi nanoparticles**

The resulting PtNi and Ga-PtNi nanoparticles were dispersed in ethanol with 3 hours of sonication. The dispersions were then drop-cast onto TEM grids with 3-4 nm thick carbon membranes and dried in air at room temperature. The TEM grid was mounted onto a GCE) using a Teflon cap, to ensure proper electrical contact. Electrochemical tests were conducted in a three-electrode system using a potentiostat (PGSTAT302N, Metrohm Autolab). The TEM grid on the GCE served as the working electrode, while a platinum mesh and a saturated calomel electrode (Hg/Hg$_2$Cl$_2$) were employed as the counter electrode and reference electrode, respectively (Supplementary Fig. 22). The 0.1 M HClO$_4$ electrolyte (with a pH of 1.0 ± 0.1) was prepared by mixing perchloric acid and deionized water, and the electrolyte was utilized for all electrochemical characterizations. The prepared electrolyte was utilized within 24 hours. According to the Nernst equation ($E_{RHE} = E_{SCE} + 0.241 + 0.0592 \times pH$), all measured potentials from the SCE were converted to the RHE.

*Catalyst Activation*

Prior to cyclic voltammetry and potential cycling, each TEM grid was directly mounted onto a GCE using a Teflon cap. The TEM grids underwent activation in an argon (Ar)-saturated 0.1 M HClO$_4$ solution in a potential range from 0.05 to 1.2 V vs. RHE with a scan rate of 50 mV s$^{-1}$, which was continued until the current became stabilized.

*Cyclic Voltammetry*

To confirm the electrical contact between the GCE and TEM grid, cyclic voltammetry was conducted in an argon (Ar)-saturated 0.1 M HClO$_4$ solution in a potential range from 0.05 to 1.2 V vs. RHE with a scan rate of 50 mV s$^{-1}$ (Supplementary Fig. 23). Identical cyclic voltammetry was performed using a bare GCE as a reference.

*Potential Cycling*

For AET analysis at various numbers of cycles, at least one TEM grid was prepared for each cycle and doping configuration (pristine and 12k cycles for undoped PtNi, and pristine, 4k, 8k, and 12k cycles for Ga-PtNi). The TEM grid on the GCE was subjected to potential cycling between 0.6 and 1.0 V vs. RHE in an oxygen ($O_2$)-saturated 0.1 M $HClO_4$ solution at a scan rate of 100 mV s$^{-1}$, serving as an accelerated stress test. After the potential cycling, the samples were annealed in a vacuum at 150 °C for 10 hours to reduce hydrocarbon contamination.

**ORR activity and durability evaluation of PtNi/C and Ga-PtNi/C catalysts**

The ORR activity and durability of the catalysts were evaluated for PtNi/C and Ga-PtNi/C. The catalyst inks were prepared by mixing 5 mg of catalyst powder with 6 μL of 20 wt% Nafion solution, 1870.5 μL of deionized water and 623.5 μL of 2-propanol, followed by ultrasonication for 3 hours. The resulting inks were drop-cast onto a GCE (with a geometric area of 0.1963 cm$^2$) with a Pt loading of 10 μg$_{Pt}$ cm$_{geo}^{-2}$. Electrochemical measurements were conducted in a three-electrode system, with all potentials calibrated to the RHE, as described above. ORR polarization curves were recorded in an oxygen-saturated 0.1 M $HClO_4$ solution over the potential range from 0.05 to 1.1 V vs. RHE with a scan rate of 20 mV s$^{-1}$ at a rotating speed of 1600 rpm (Supplementary Fig. 17a,b). Cyclic voltammetry was performed in an argon-saturated 0.1 M $HClO_4$ solution between 0.05 and 1.2 V vs. RHE with a scan rate of 50 mV s$^{-1}$ (Supplementary Fig. 17c,d). Note that the ORR polarization curves and cyclic voltammograms were also recorded after 12k potential cycles conducted under the same conditions used for AET observation.

The mass activities (MAs) were determined by ORR polarization curves using the Koutecky-Levich equation:

$$\frac{1}{J} = \frac{1}{J_d} + \frac{1}{J_k}. \qquad (1)$$

In this equation, $J$ is the measured current density. $J_d$ and $J_k$ are the limiting current by diffusion and the kinetic current density, respectively, where the $J_d$ is defined by the Levich equation:

$$J_d = 0.62 n F D^{2/3} v^{-1/6} \omega^{1/2} C_{O_2}, \qquad (2)$$

where $n$ is the number of electrons transferred, F is the Faraday constant (96485 C mol$^{-1}$), $D$ is the diffusion coefficient of oxygen in 0.1 M $HClO_4$ solution (1.93 × 10$^{-5}$ cm$^2$ s$^{-1}$), $v$ is the kinematic viscosity of the electrolyte (1.01 × 10$^{-2}$ cm$^2$ s$^{-1}$), $\omega$ is the angular frequency of rotation of the RDE ($\omega = 2\pi f / 60$), $f$ is the rotating speed of the RDE in rpm, and $C_{O_2}$ is the concentration of molecular oxygen in 0.1 M $HClO_4$ solution (1.26 × 10$^{-6}$ mol cm$^{-3}$).

The MAs are calculated by normalization of the kinetic current density with the loading amount of Pt. At a potential of 0.9 V vs. RHE, the initial MAs were determined to be 0.81 A mg$_{Pt}^{-1}$ for PtNi/C and 1.78 A mg$_{Pt}^{-1}$ for Ga-PtNi/C, which decreased to 0.41 (a 49% reduction) and 1.47 A mg$_{Pt}^{-1}$ (a 17% reduction), respectively, after 12k cycling (Supplementary Fig. 18a).

The electrochemically active surface areas (ECSAs) were calculated from the cyclic voltammograms by integrating the hydrogen oxidation charge from 0.05 to 0.4 V vs. RHE. The ECSA were obtained by the following relationship:

$$\text{ECSA} = \frac{Q_{H-desorption}}{Q_{H-monolayer}\, L_{Pt}}, \qquad (3)$$

where $Q_{H-desorption}$ is the hydrogen oxidation charge normalized with a scan rate of the cyclic voltammograms, $Q_{H-monolayer}$ is the conversion factor, assuming a hydrogen oxidation charge of 210 μC cm$^{-2}$ for a monolayer of hydrogen adsorbed on Pt, and $L_{Pt}$ is the loading amount of Pt.

The ECSAs, estimated from the hydrogen oxidation charge, were 23.7 m$^2$ g$_{Pt}^{-1}$ for PtNi/C and 19.1 m$^2$ g$_{Pt}^{-1}$ for Ga-PtNi/C. After 12k potential cycles, ECSAs remained relatively stable at 23.0 and 22.9 m$^2$ g$_{Pt}^{-1}$, respectively (Supplementary Fig. 18b). The specific activities were obtained from the MAs by normalizing with the ECSAs. The electrochemical evaluation of PtNi/C and Ga-PtNi/C was recorded only once, and the error bar of the derived characteristics cannot be defined.

**STEM data acquisition**

The tomographic tilt series were acquired using three different microscopes: an FEI Titan Themis Z at the Institute of Next-generation Semiconductor convergence Technology (INST), the TEAM 0.5 microscope equipped with the TEAM stage at the National Center for Electron Microscopy (NCEM), and a Thermo Fisher Scientific Spectra Ultra TEM at KARA. From the tilt series datasets, atomic resolution electron tomograms were obtained for nanoparticles of 2 PtNi pristine, 6 PtNi 12k, 3 Ga-PtNi pristine, 2 Ga-PtNi 4k, 2 Ga-PtNi 8k, and 2 Ga-PtNi 12k cycles (see Supplementary Tables 1 and 2). Note that various cycling was performed on different nanoparticles, not on the same nanoparticle. The images for each tilt series were obtained using the ADF-STEM mode with 200 kV or 300 kV acceleration voltage. To correct the effect of stage drift during measurement, three or four consecutive images of 1024 × 1024 pixels were collected using a dwell time of 3 μs at each tilt angle (see Supplementary Tables 1 and 2 for detailed microscope parameters). Each tilt series was measured using a total electron dose in the range of $2.1 \times 10^5$ $e$ Å$^{-2}$ to $4.0 \times 10^5$ $e$ Å$^{-2}$. To ensure minimal structural changes induced by the electron beam, zero-degree projections were measured three times: at the beginning, in the middle, and at the end of the tilt series acquisition for comparison (see Supplementary Figs. 9 and 10). As shown in Supplementary Figs. 9 and 10, no substantial structural changes were observed, indicating that the PtNi nanoparticles remain stable within the given total electron dose range.

**Image post-processing**

We performed image post-processing for the tilt series using a series of procedures, including drift correction, scan distortion correction, image denoising[63], background subtraction, and tilt-series alignment based on center-of-mass and common-line alignment, following the methods described in previous works[37–39,48,64–66].

(I) Drift and scan distortion correction: We performed drift correction on the images at each tilt angle by estimating the linear stage drift from three or four consecutively acquired ADF-STEM images and compensating for it using an affine transform. Subsequently, the scan distortion of the images was corrected using a scan distortion matrix estimated from images of a single-crystal silicon (110) standard sample. The three or four consecutive images, after drift and scan distortion correction, were then averaged to create an experimental image for each tilt angle.

(II) Image denoising: To remove the Gaussian-Poisson mixed noise in the ADF-STEM images, we applied the block-matching and 3D filtering (BM3D) algorithm[63] using Gaussian-Poisson noise parameters estimated from the experimental tilt series.

(III) Tilt-series alignment: To remove the background signal within the nanoparticles, we defined a 2D mask slightly larger than the boundary of the nanoparticles for each projection image. We then solved the Dirichlet boundary value problem of the discrete Laplace's equation to determine the background signal, which was subsequently subtracted from each denoised image. Afterward, each tilt series was aligned with sub-pixel accuracy using the center-of-mass[36] and common-line alignment[38] methods.

**Generation of training/test datasets for the neural network-assisted AET**

Input and target datasets were prepared for training, validation, and testing of the neural networks following the previously established procedures[39,42]. A total of 4,800 simulated tomograms were generated, with 4,000 for training, 400 for validation, and 400 for testing. The generation of input data followed a four-step process.

(1) We first created a blank 3D volume with a size of 384 × 384 × 384 voxels and a voxel size of 0.329 Å. Within this blank volume, we constructed a random 3D shape with a volume ranging from 865,700 to 1,121,000 Å$^3$, and placed a randomly oriented fcc atomic structure of a 3.84 Å lattice constant inside the random 3D shape. The atomic structures included point defects, with percentages randomly chosen between 0% and 0.5%, and random spatial displacements of about 20 pm root-mean-square deviation (RMSD). The generated atomic positions were then randomly assigned as either Pt or Ni atoms, maintaining a Pt-to-Ni ratio of 1:1.

(2) The sharp 3D atomic potentials at the atomic positions were calculated[39,42,67], and the resulting sharp potentials were convolved with Gaussian kernels to simulate broadening effects, including thermal vibrations. The standard deviations ($\sigma$) of the Gaussian kernels for Ni and Pt atoms were randomly chosen from a Gaussian distribution with a mean of 0.60 Å for Ni and 0.70 Å for Pt, and standard deviation of 0.11 Å for both. To account for the intensity ratio of Pt and Ni atoms in experimentally reconstructed tomograms, a weighting factor of 0.7 was applied to the Pt atomic potential.

(3) 41 forward-projections of the calculated 3D volumes for tilt angles ranging from −65° to +65° were generated as a tilt series for each volume. The projection size was 384 × 384 pixels with a pixel size of 0.329 Å. To mimic the real experiment conditions, Poisson noise and random tilt angle errors up to ±0.3° were added.

(4) 3D tomograms were reconstructed from the generated tilt series and the tilt angles using the GENFIRE algorithm[47]. The GENFIRE parameters of the fast Fourier transform interpolation method with 100 iterations, an oversampling ratio of 2, and an interpolation radius of 0.3 pixels were used for the reconstructions. To normalize the intensity of the 3D volumes, we first calculated the average intensity of the 3 × 3 × 3 voxels at all the Pt atomic positions. Then, we divided the reconstructed volume by this averaged intensity.

For generating target data (ground truth), we followed steps (1) and (2) of the input data generation procedure. The only difference was that Gaussian kernels with the standard deviations (σ) of 0.60 Å for Ni and 0.70 Å for Pt were applied to the calculated sharp potentials. For intensity normalization, we also divided the generated target volume by the average intensity calculated in step (4).

**The framework of deep learning neural network and training**

The deep learning neural network used in this study follows the framework suggested in our previous work[42]. The primary difference is the increased depth; we utilized a neural network that is two layers deeper than that of the previous study[42] (see Supplementary Fig. 24). As shown in Supplementary Fig. 24, both the input and output data sizes were 384 × 384 × 384 voxels. The main building blocks of the network include 3 × 3 × 3 convolutions with a stride of 2 for down-sampling, 2 × 2 × 2 max-pooling for down-sampling, 3 × 3 × 3 transposed convolutions with a stride of 2 for up-sampling, and two 3 × 3 × 3 convolutions with a stride of 1. To prevent overfitting, we employed the dropout method[68]. Two activation functions were used: the Leaky Rectified Linear Unit (Leaky ReLU)[69] with a leakage coefficient of 0.2 for all layers except the final output layer, and the Rectified Linear Unit (ReLU)[70] for the final output layer. The mean squared error was used as the loss function, and the Adam optimizer[71] was employed with a learning rate of $1 \times 10^{-3}$. Supplementary Figure 25 displays the learning curve for the training and validation sets, showing no signs of overfitting or divergence throughout the training process. The trained neural network model at the 30th epoch was used for the augmentation of experimental tomograms.

**3D reconstruction and neural network-based volume data augmentation**

After image post-processing, 3D tomograms were reconstructed from the post-processed tilt series using the GENFIRE algorithm[47]. To enhance the quality of the tomograms, we applied angular refinement[47] implemented in the GENFIRE algorithm, as well as in-plane rotational and translational re-alignment[39]. After the corrections, we reconstructed the final 3D tomograms using the following GENFIRE parameters: discrete Fourier transform interpolation method, number of iterations of 1000, an oversampling ratio of 4, and an interpolation radius of 0.1 pixels (Supplementary Tables 1 and 2). However, the tomograms suffered from artifacts caused by data imperfections due to sparse sampling and the missing wedge problem. To address this, we augmented the tomograms using the deep learning-based neural network as described above[39,42]. To normalize the experimental tomograms for input into the trained neural network for tomogram augmentation, we calculated the average integrated intensity, $I_{ave}$, of the approximately $1.0 \times 1.0 \times 1.0$ Å$^3$ ($3 \times 3 \times 3$ voxels) volumes around the Pt atom positions in the experimental tomograms. Several normalization factors were tested, and dividing the raw experimental tomograms by $\frac{I_{ave}}{1.3}$ demonstrated the best performance for neural network-based volume augmentation, specifically in terms of atom tracing as discussed in the next section. Therefore, this normalization factor was consistently applied to all experimental tomograms before applying the data augmentation.

**Atom tracing and species classification**

For 3D tomograms before and after neural network-based volume data augmentation, we determined the 3D atomic coordinates using the following atom tracing and species classification processes[38,40,48].
(i) The 3D local maxima positions in the tomograms were identified and ranked by peak intensity in descending order. Starting with the highest intensity, a volume of $5 \times 5 \times 5$ voxels ($1.6 \times 1.6 \times 1.6$ to $1.8 \times 1.8 \times 1.8$ Å$^3$, depending on voxel size) centered on each local maximum was extracted, and a 3D Gaussian function was fitted to the cropped volume to determine the precise peak position. The identified peak position was added to a list of potential atom positions only if it satisfied a minimum distance of 2.0 Å from any previously listed fitted positions. This procedure was repeated for all 3D local maxima, resulting in a list of potential atom positions.
(ii) A *k*-means clustering algorithm[38,40,48] was used to classify the potential atom positions into three types of atoms (non-atom, Ni, and Pt). Atoms classified as Ni or Pt located outside the 3D boundary of the nanoparticle, as defined by the Otsu threshold, were further reclassified as non-atom. Note that due to the similarity in atomic numbers between Ga and Ni, they cannot be distinguished in AET, resulting in Ga being counted as Ni; however, since the compositional change of Ni in Ga-PtNi during cycling varies by more than 7 percentage points (from 42% to 35%), the 1.5% Ga doping does not affect our analysis regarding the dynamics of Ni fractions much.
(iii) To fully determine the atomic structures that were not identified in the previous steps, we applied an additional atom tracing process[40,48]. Since the PtNi nanoparticles have an fcc crystal structure, the tomogram was first rotated to align with the crystallographic direction. The rotated 3D tomogram was sliced along the [001] direction for each atomic layer, with the slicing positions corresponding to the peak positions in the histogram of the atom position components along the [001] direction. We identified the positions of 2D local maxima, along with their corresponding intensities, for all slices. Starting with the highest intensity, the 3D Gaussian fitting procedure was performed using the 2D local maxima positions as initial estimates. The size of the cropped volume for the 3D Gaussian fitting was adjusted by varying the side length from 3 to 7 pixels. The best-fit position, determined from the volume size yielding the smallest mean squared residuals, was added to the traced atom list, provided that it satisfied a minimum distance of 2 Å from neighboring atoms. To finalize the 3D atomic structures and their chemical compositions, we repeated the process described in (ii), and as a result, obtained the full 3D atomic structures of 8 PtNi and 9 Ga-PtNi nanoparticles.

(iv) To verify the consistency between the determined atomic structures and the measured projections, we calculated the R-factors by comparing the experimentally obtained tilt series with the simulated tilt series generated from the forward projections of the final 3D atomic models. The average R-factor for all nanoparticles was determined to be 0.09 ± 0.01 (Supplementary Tables 1 and 2), a value considered acceptable in the crystallography community, indicating consistency between the two sets of images.

**Assignment of experimental 3D atomic positions to ideal fcc lattices**

The 3D atomic positions were assigned to ideal fcc lattice sites through the following procedures[42,40].

(a) A Pt atom nearest to the average position of all the 3D atomic coordinates was selected as the origin of an ideal fcc lattice.

(b) Subsequently, the nearest fcc sites to this origin atom position were calculated using the initial fcc lattice vectors, with a lattice constant of 3.9 Å. For each calculated fcc site, if an atom was located within 25% of the nearest neighbor distance of the fcc lattice, the atom was assigned to that corresponding fcc lattice site.

(c) The nearest neighbor search was repeated for all newly assigned fcc lattice sites, and this process was continued until no additional atoms could be assigned to the lattice.

(d) New fcc lattice vectors were fitted to the assigned atoms by changing the translation, 3D rotation, and lattice constant parameters, aiming to minimize the discrepancy between the 3D atomic positions and their corresponding lattice sites in the fitted fcc lattice.

(e) The processes from (a) to (d) were performed iteratively until the fitted lattice vectors remained unchanged.

**Precision estimation using STEM multislice simulation**

A precision estimation was conducted to evaluate the reliability of the 3D atomic models of PtNi and Ga-PtNi nanoparticles. Projection images were generated from the finalized 3D atomic models at experimental tilt angles using an efficient multislice simulation, specifically the PRISM simulation[72,73], with an interpolation factor of 2, a slice thickness of 2 Å, sixteen frozen phonon configurations, and aberration values of −352.8 nm $C_3$ and 1.5 mm $C_5$ for nanoparticles imaged at INST, 0 nm $C_3$ and 5 mm $C_5$ for nanoparticles imaged at NCEM, and −521 nm $C_3$ and 0.58 mm $C_5$ for nanoparticles imaged at KARA. Other microscope parameters, such as electron acceleration voltages, convergence semi-angles, and detector inner and outer semi-angles were adjusted to match the experimental parameters provided in Supplementary Tables 1 and 2. To account for the influence of electron probe size and other incoherent effects, a Gaussian kernel with an optimized standard deviation was applied to each PRISM-simulated image (see Supplementary Tables 1 and 2).

We applied the GENFIRE algorithm[47] to reconstruct the 3D tomograms from the PRISM-simulated tilt series, using the same parameters described above. The 3D atomic structures of Pt and Ni atoms were subsequently determined from these tomograms using previously described methods (see above for more details on atom tracing and species classification).

To match the Pt-to-Ni intensity ratio of the tomograms from the multislice simulation with those from the experiment, we performed additional processing. We first obtained the average Pt ($I_{Pt}^{exp}$) and Ni intensity ($I_{Ni}^{exp}$) by averaging the intensities within approximately $1.0 \times 1.0 \times 1.0$ Å$^3$ ($3 \times 3 \times 3$ voxels) volumes surrounding the atom positions in the experimental tomograms of each nanoparticle. Based on these values, the Pt-to-Ni intensity ratio, $I_{Pt}^{exp}/I_{Ni}^{exp}$, was calculated, and the average Pt-to-Ni intensity ratio across all nanoparticles under our experimental conditions was determined to be 1.6. To ensure that the Pt-to-Ni intensity ratio of tomograms from PRISM simulation matches the ratio determined from the experimental data, a constant background value, $I_{const}$, was added to the 3D tomogram obtained from the PRISM simulation, such that the condition, $\frac{I_{Pt}^{sim}+I_{const}}{I_{Ni}^{sim}+I_{const}} = \frac{I_{Pt}^{exp}}{I_{Ni}^{exp}} = 1.6$, is satisfied. Here, $I_{Pt}^{sim}$ and $I_{Ni}^{sim}$ represent the intensity values of Pt and Ni atoms, respectively, obtained from the 3D tomogram from the multislice simulation.

Subsequently, the neural network-based volume data augmentation process was performed as previously described, and the 3D atomic structures were finally determined. To compare the atomic structures obtained from experiments with those from PRISM simulations, we calculated the distances between atoms common to both the experimental and PRISM-simulated structures. Atom pairs with distances less than a specified threshold (half of the first nearest-neighbor distance in the ideal fitted fcc lattice, 1.4 Å) were classified as common atom pairs. The analysis revealed that, on average, 94.2 ± 2.6% of atoms were successfully identified in the simulation. Additionally, the average root-mean-square deviation (RMSD) of all common atom pairs across all nanoparticles (i.e., the precision of the atomic coordinates; see Refs.[37,38,64]) was determined to be 31.7 ± 5.2 pm (see Supplementary Tables 1 and 2).

Furthermore, based on the precision of atomic coordinates for each nanoparticle, the precision of the average strain values for surface Pt atoms was calculated through standard error propagation for each 3D atomic structure (Supplementary Tables 7 and 8). The precision ranges from 0.06% to 0.12%, with a mean value slightly below 0.1%.

**Calculation of sphericity**

To determine how closely each given nanoparticle's shape resembles that of a perfect sphere, we calculated the sphericity[74], defined as $\pi^{1/3}(6V)^{2/3}/A$, where A and V represent the surface area and the volume of the nanoparticle, respectively. The surface area and volume of the nanoparticle were calculated using the alpha-shape algorithm[75], with the lattice constant obtained from global fcc fitting used as the alpha radius.

**Chemical short-range order parameter (CSRO)**

To understand the local chemistry of each atom, we calculated the CSRO, which is a pairwise multicomponent short-range order parameter[76,77]. The CSRO between Pt atoms and their nearest neighbor Ni atoms ($\alpha_{PtNi}$) and the CSRO between Ni atoms and their nearest neighbor Pt atoms ($\alpha_{NiPt}$) are given by

$$\begin{cases} \alpha_{PtNi} = 1 - \frac{p_{PtNi}}{C_{Ni}} \\ \alpha_{NiPt} = 1 - \frac{p_{NiPt}}{C_{Pt}} \end{cases}, \qquad (4)$$

where $p_{PtNi}$ is the Ni atom fraction among the 1$^{st}$ nearest neighbors of Pt atoms, $C_{Ni}$ is the Ni atom fraction in the entire nanoparticle, $p_{NiPt}$ is the Pt atom fraction among the 1$^{st}$ nearest neighbors of Ni atoms, $C_{Pt}$ is the Pt atom fraction in the entire nanoparticle. A CSRO value near 0 indicates that the local region around a given atom reflects the chemical composition of the entire nanoparticle. Values above 0 suggest local chemical segregation, while values below 0 denote a higher degree of alloying than the global composition of the nanoparticle. Note that the 1$^{st}$ nearest neighbor is defined as a set of atoms within a cutoff distance determined as the midpoint between the 1$^{st}$ and 2$^{nd}$ nearest neighbors, based on the lattice constant obtained from the global fcc fitting (see 'Assignment of experimental 3D atomic positions to ideal fcc lattices' section of Methods).

**Defining the surface atoms and calculating the distance from the surface**

Surface atoms were defined as the atoms located outside a 3D mask that was 6 Å smaller than the nanoparticle boundary determined during the non-atom classification step described above. The distance from the surface for each atom was defined as the shortest distance to the nearest surface atom. Note that the criterion of 6 Å was determined to ensure that the surface atoms uniformly cover the entire nanoparticle.

**Calculation of local volumetric strain**

To obtain local strain, we performed local fcc lattice fitting using the following procedure. First, we identified a set of 2$^{nd}$ nearest neighbor atoms centered around each atom based on the globally fitted fcc lattice. Using the translation, 3D rotation, and lattice constant parameters of the globally fitted lattice as the initial condition, we performed a similar fcc lattice fitting for this set of atoms by minimizing the error between the measured atomic positions and the corresponding lattice sites. Through these determined local fcc lattice vectors, we calculated the lattice constant for each atom, which we defined as the local lattice constant ($a_{local}$). Using the local lattice constant, the local volumetric strain ($S$) for each atom was calculated using the following equation:

$$S = \left(\frac{a_{\text{local}}}{a_{\text{ref}}} - 1\right). \qquad (5)$$

Here, $a_{\text{ref}}$ and $a_{\text{local}}$ represent the lattice constant of bulk Pt[56] (3.912 Å) and the local lattice constant of each atom, respectively. The positive and negative signs of the local volumetric strain indicate tensile and compressive strain, respectively.

**Assignment of the facet indices**

To assign facet indices for each surface atom, we first calculated the normal vector using the discrete Laplace-Beltrami operator via cotangent discretization. For each surface atom, we formed a set of atoms within a cutoff distance defined as $\frac{1}{2}\left(\frac{a_{\text{global}}}{\sqrt{2}} + a_{\text{global}}\right)$, where $a_{\text{global}}$ is the lattice constant obtained from the global fcc fitting. We then computed the convex hull of this atom set and calculated the mean curvature normal operator using the cotangent discretization method[50]. The mean curvature normal operator, evaluated at the given surface atom position, was normalized and defined as the normal vector of the atom. We further applied 3D Gaussian kernel averaging to the normal vectors with a standard deviation of $\frac{a_{\text{global}}}{\sqrt{2}}$, equivalent to the 1st nearest neighbor distance. For the kernel averaged normal vector for each surface atom, we calculated dot products with the normal vectors of three crystallographic facet families ({100}, {110}, and {111}) and assigned it to the facet with the largest dot product value (Supplementary Movies 1, 2 and Supplementary Figs. 26, 27).

**Calculation of the ORR activity**

To calculate the ORR activity for alloy nanocatalysts, we combined the strain effect and ligand effect with the GCN.

Typically, the GCN[20] for atom $i$ can be expressed as

$$\text{GCN}(i) = \sum_{j=1}^{n_i} \frac{\text{CN}(j)}{\text{CN}_{\max}}. \qquad (6)$$

Here, $n_i$ represents the coordination number for atom $i$, CN($j$) denotes the coordination number of the 1st nearest neighbor atom $j$ of atom $i$, and CN$_{\max}$ is the number of 1st nearest neighbors in the bulk (12 for fcc).

By incorporating a prefactor that accounts for the strain effect[55] and a correction factor for Ni (1.59) accounting for the ligand effect (estimated from density functional theory calculation[57]), the extended generalized coordination number (so-called alloy-sensitive-GCN; ASGCN[57]) for atom $i$ can be expressed as

$$\text{ASGCN}(i) = \frac{1}{1+S(i)}(1.59 \times \text{GCN}(i)_{\text{Ni}} + \text{GCN}(i)_{\text{Pt}}), \qquad (7)$$

where GCN($i$)$_{\text{Ni}}$ and GCN($i$)$_{\text{Pt}}$ represent the contributions of the Ni and Pt atoms of the 1st nearest neighbors to the GCN value, respectively, therefore $[\text{GCN}(i)_{\text{Ni}} + \text{GCN}(i)_{\text{Pt}}] = \text{GCN}(i)$. Since alloy-sensitive GCN exhibits a volcano relationship with ORR activity $[\ln(j/j_{Pt(111)})]$[57], we can calculate ORR activity at the single-atom level by determining the ASGCN for each Pt surface atom.

However, in the case of the ASGCN, simulations do not accurately represent actual experiments. Specifically, when estimating the ligand effect of Ni, simulations assume that the surface atoms are composed solely of Pt and the Ni-Pt alloy can form only in the subsurface. In contrast, actual nanocatalysts do not have a perfectly homogeneous Pt surface; instead, they exhibit local chemical heterogeneity.

Therefore, the ligand effects of subsurface and surface Ni should be considered separately. In general, many simulations account only for the ligand effect of subsurface Ni[57,78], leading us to assume that the ligand effect of surface Ni can be considered negligible compared to that of subsurface Ni. Consequently, the alloy-sensitive GCN can be revised to the modified alloy-sensitive GCN (MASGCN), as shown in Eq. (8):

$$\text{MASGCN}(i) = \frac{1}{1+S(i)} \left( 1.59 \times \text{GCN}(i)_{\text{Ni(subsurface)}} + \text{GCN}(i)_{\text{Ni(surface)}} + \text{GCN}(i)_{\text{Pt}} \right). \tag{8}$$

The ORR calculation results in Fig. 4 are calculated based on the MASGCN above.

**EDS experiment**

An EDS analysis was conducted using a Titan Double Cs-corrected TEM (Titan Cubed G2 60-300, FEI) operated at 300 kV at KARA. The specimen was prepared by depositing a solution of the synthesized PtNi or Ga-PtNi nanoparticles dispersed in water onto multiple carbon membrane grids (3-4 nm membrane thickness), followed by vacuum annealing at 150 °C for 24 h. Potential cycling was selectively conducted for these grids to prepare at least one TEM grid for each cycle and doping configuration (pristine and 12k cycles for PtNi, and pristine, 4k, 8k, and 12k cycles for Ga-PtNi). The EDS data were collected using a Super-X EDS detector operated in STEM mode under the following conditions: acceleration voltage of 300 kV, screen current of approximately 150 pA, pixel dwell time of 10 μs, and an image resolution of 512 × 512 pixels. A total of 13, 13, 15, 15, 15, and 16 EDS elemental mapping images were acquired for PtNi-pristine, PtNi-12k, Ga-PtNi-pristine, Ga-PtNi-4k, Ga-PtNi-8k, and Ga-PtNi-12k nanoparticles, respectively.

The composition of PtNi and Ga-PtNi nanoparticles from the EDS mappings is summarized in Supplementary Table 3. Before cycling, PtNi (60 at% Pt) and Ga-PtNi (61 at% Pt) show a consistent composition. Given that the error margin for EDS-based composition quantification is approximately 4 at%[79,80], the Pt content ranges for pristine PtNi (59-63 at%) and Ga-PtNi (56-60 at%) obtained via AET align fully with the EDS results. This consistency further confirms that there is no meaningful difference in the pristine compositions of PtNi and Ga-PtNi nanoparticles.

**Calculation of the number of nanoparticles on the TEM grid**

We measured the loading of nanoparticles on our TEM grid as follows. First, we prepared TEM specimens of PtNi and Ga-PtNi nanoparticles under the same conditions as those used for potential cycling and AET experiments, as described above. To quantify the nanoparticle loading, we acquired 36 images of PtNi nanoparticles and 22 images of Ga-PtNi nanoparticles at low-magnification (Supplementary Fig. 28a for PtNi and Supplementary Fig. 29a for Ga-PtNi). The images were obtained using a Titan Double Cs-corrected TEM (Titan Cubed G2 60-300, FEI) with ADF-STEM mode. All images were captured under the following microscope parameters: an acceleration voltage of 300 kV, detector inner and outer semi-angles of 38 mrad and 200 mrad, respectively, a convergence beam semi-angle of 18.0 mrad, and a beam current of approximately 150 pA. The image size was $1024 \times 1024$ pixels, with a pixel dwell time of 4 μs. The pixel size was 13.5 nm for PtNi images and 18.5 nm for Ga-PtNi images. We calculated the area occupied by the nanoparticles in the images using the Otsu thresholding method[81] to separate nanoparticle regions from the background. Additionally, to estimate the size of individual nanoparticles, we acquired 23 ADF-STEM images for PtNi nanoparticles and 14 images for Ga-PtNi nanoparticles at high-magnification, in which single nanoparticles can be clearly distinguished. The Otsu thresholding method was applied to these images again, and dividing the area of particle regions by the number of nanoparticles yielded the average area per nanoparticle of 74.8 nm$^2$ for PtNi and 33.8 nm$^2$ for Ga-PtNi. To estimate the number of nanoparticles in each image, the total area occupied by nanoparticles in the low-magnification images was divided by the average area of PtNi and Ga-PtNi single nanoparticles. The estimated numbers are plotted in Supplementary Fig. 28b for PtNi and Supplementary Fig. 29b for Ga-PtNi. Considering the total field of view, the sample density was determined to be 540 μm$^{-2}$ for PtNi and 1140 μm$^{-2}$ for Ga-PtNi. Finally, multiplying the calculated density by the total TEM grid area yielded an estimated loading of nanoparticles on the grid: $3.94 \times 10^9$ for PtNi and $8.32 \times 10^9$ for Ga-PtNi.

## Data availability

All of our experimental data, tomographic reconstructions, determined atomic structures, and ORR results are posted on a public website (https://mdail.kaist.ac.kr/NiPt_dynamics), and they can also be accessed through an open repository (https://doi.org/10.5281/zenodo.15201029).

## Code availability

Source codes are posted on a public website (https://mdail.kaist.ac.kr/NiPt_dynamics), and they can also be accessed through an open repository (https://doi.org/10.5281/zenodo.15201029).

## Acknowledgements

This research was supported by the National Research Foundation of Korea (NRF) Grants funded by the Korean Government (MSIT) (No. RS-2023-00208179). Y.Y. also acknowledges the support from the KAIST singularity professor program. Part of the STEM experiments were conducted using a Spectra Ultra TEM at the KAIST Analysis Center for Research Advancement (KARA) and a Titan Themis Z (FEI) equipment at the DGIST Institute of Next-generation Semiconductor convergence Technology (INST). Excellent support by Jin-Seok Choi, Taehoon Cheon, and the staff of KARA and INST is gratefully acknowledged. The experiment utilizing the TEAM 0.5 microscope was performed at the Molecular Foundry, which is supported by the Office of Science, Office of Basic Energy Sciences of the US Department of Energy under contract no. DE-AC02-05CH11231. The tomography data analyses were partially supported by the KAIST Quantum Research Core Facility Center (KBSI-NFEC grant funded by Korea government MSIT, PG2022004-09). We declare that the authors utilized the ChatGPT (https://chat.openai.com/chat) for language editing purpose only, and the original manuscript texts were all written by human authors, not by artificial intelligence.

## Author Contributions

Y.Y. conceived the idea and directed the study. S.L., K.L., and E.C. synthesized the PtNi/Ga-PtNi nanoparticles and conducted the potential cycling. C.J., J.L., H.J., P.E., and Y.Y. designed and performed the electron tomography experiments. J.L. and Y.Y. designed and trained the neural network. C.J., C.O., H.J., K.L., and Y.Y. conducted the data analysis. C.J., J.L., H.J., K.L., and Y.Y. wrote the manuscript. All authors commented on the manuscript.

## Competing interests

C.J., J.L., H.J., S.L., K.L., E.C., and Y.Y. have patent application (Korea, 10-2024-0131645), which disclose 3D structure and property characterization techniques for nanocatalysts. The remaining authors declare no competing interests.

## Additional information

Correspondence and requests for materials should be addressed to E.C. (eacho@kaist.ac.kr) or Y.Y. (yongsoo.yang@kaist.ac.kr)

# Supplementary Information

# for

# Atomic-scale 3D structural dynamics and functional degradation of Pt alloy nanocatalysts during the oxygen reduction reaction

Chaehwa Jeong[1†], Juhyeok Lee[1,2,3†], Hyesung Jo[1†], KwangHo Lee[4†], SangJae Lee[4], Colin Ophus[5,6], Peter Ercius[3], EunAe Cho[4*] and Yongsoo Yang[1,7*]

[1] *Department of Physics, Korea Advanced Institute of Science and Technology (KAIST), Daejeon 34141, Republic of Korea*

[2] *Energy Geosciences Division, Lawrence Berkeley National Laboratory, Berkeley, CA 94720, USA*

[3] *National Center for Electron Microscopy, Molecular Foundry, Lawrence Berkeley National Laboratory, Berkeley, CA 94720, USA*

[4] *Department of Materials Science and Engineering, Korea Advanced Institute of Science and Technology (KAIST), Daejeon 34141, Republic of Korea*

[5] *Department of Materials Science and Engineering, Stanford University, Stanford, CA 94305, USA*

[6] *Precourt Institute for Energy, Stanford University, Stanford, 94305, USA*

[7] *Graduate School of Semiconductor Technology, School of Electrical Engineering, Korea Advanced Institute of Science and Technology (KAIST), Daejeon 34141, Republic of Korea*

[†] These authors contributed equally to this work.

*Corresponding author, email: eacho@kaist.ac.kr, yongsoo.yang@kaist.ac.kr

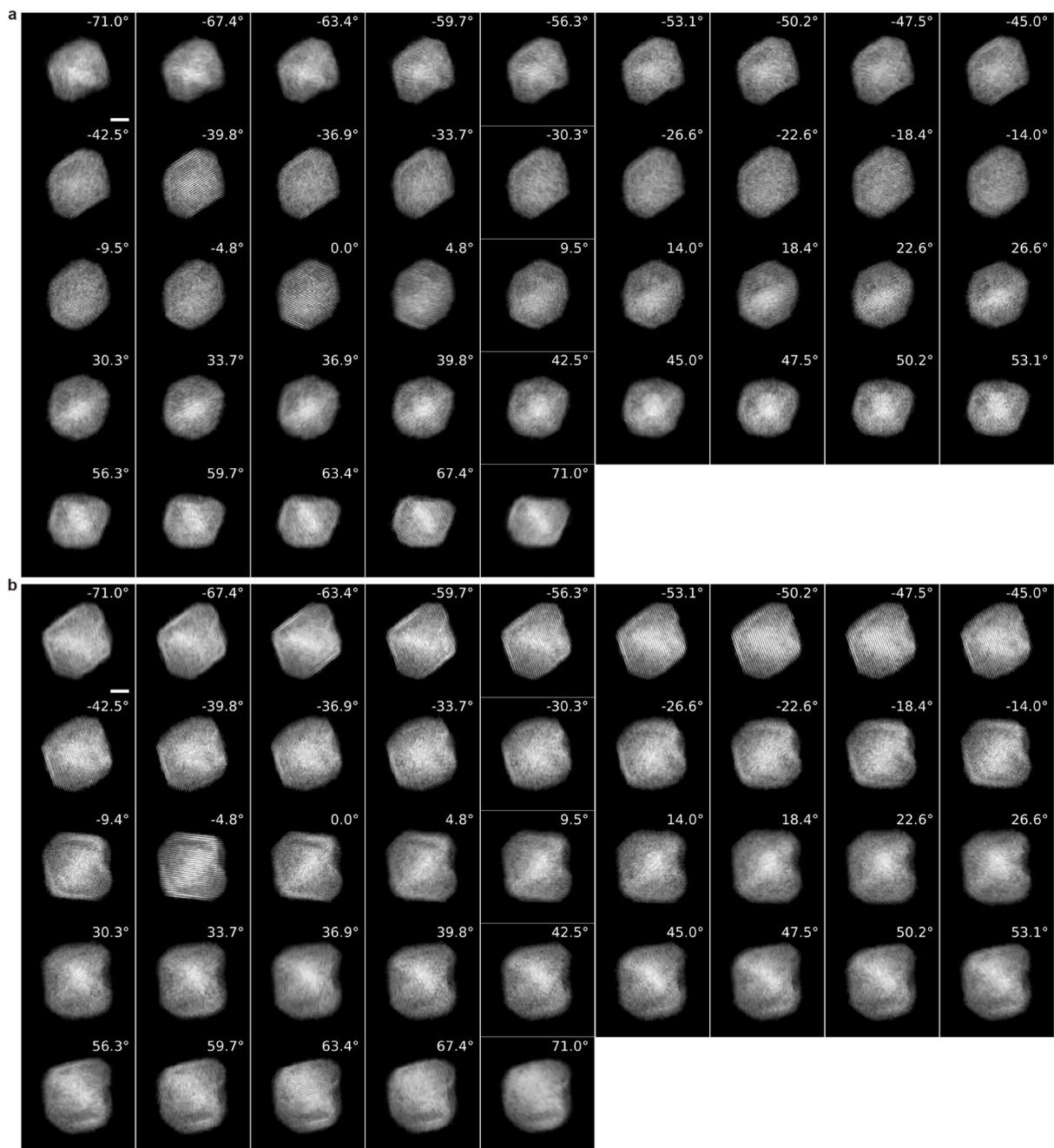

**Supplementary Figure 1 | Experimental tomographic tilt series of the pristine PtNi nanoparticles. a**, **b**, Post-processed ADF-STEM images at different tilt angles (white text in each figure), for PtNi-pristine-p1 (**a**) and PtNi-pristine-p2 (**b**). Scale bar, 2 nm.

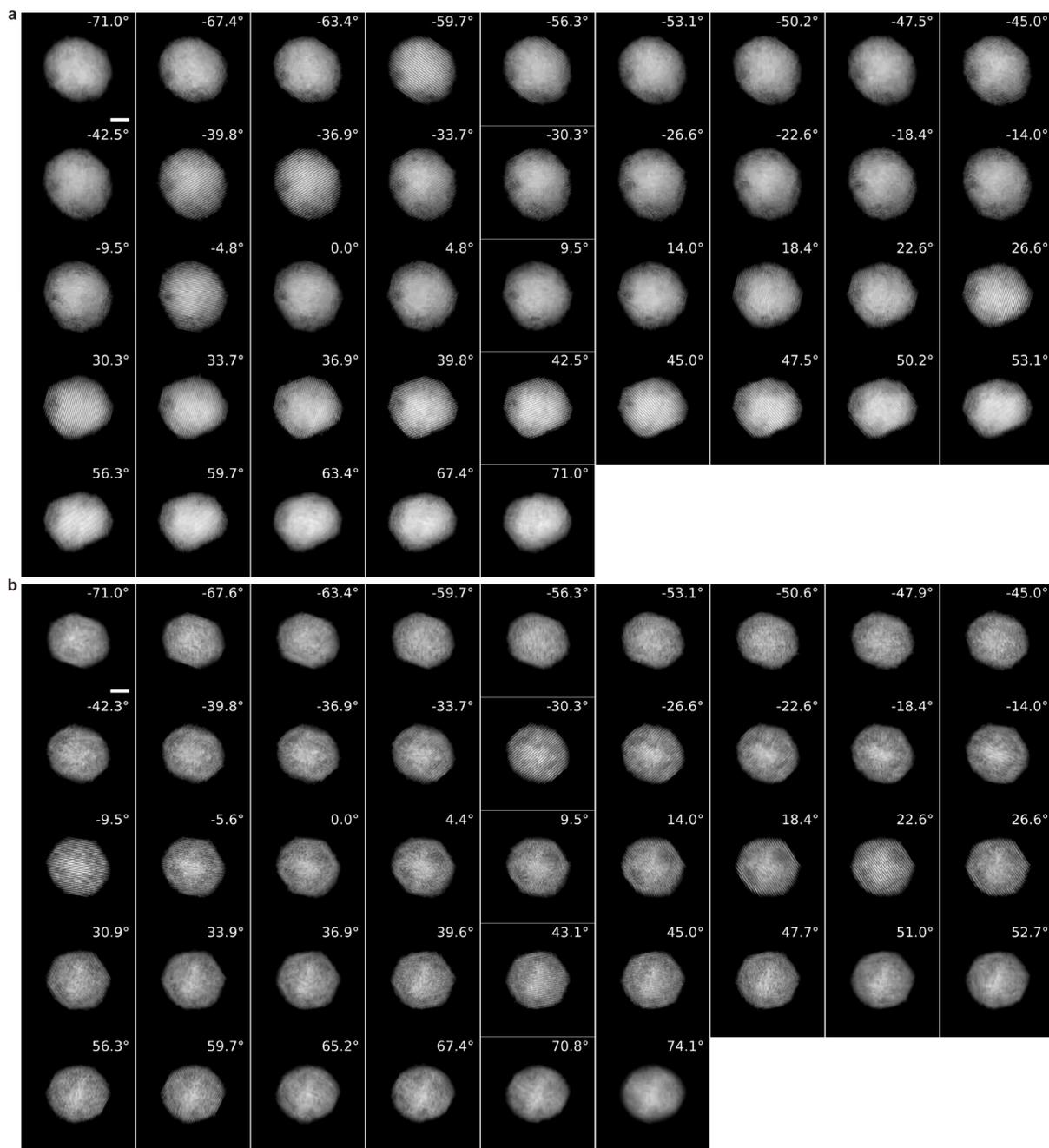

**Supplementary Figure 2 | Experimental tomographic tilt series of the 12k PtNi nanoparticles. a, b,** Post-processed ADF-STEM images at different tilt angles (white text in each figure), for PtNi-12k-p1 (**a**) and PtNi-12k-p2 (**b**). Scale bar, 2 nm.

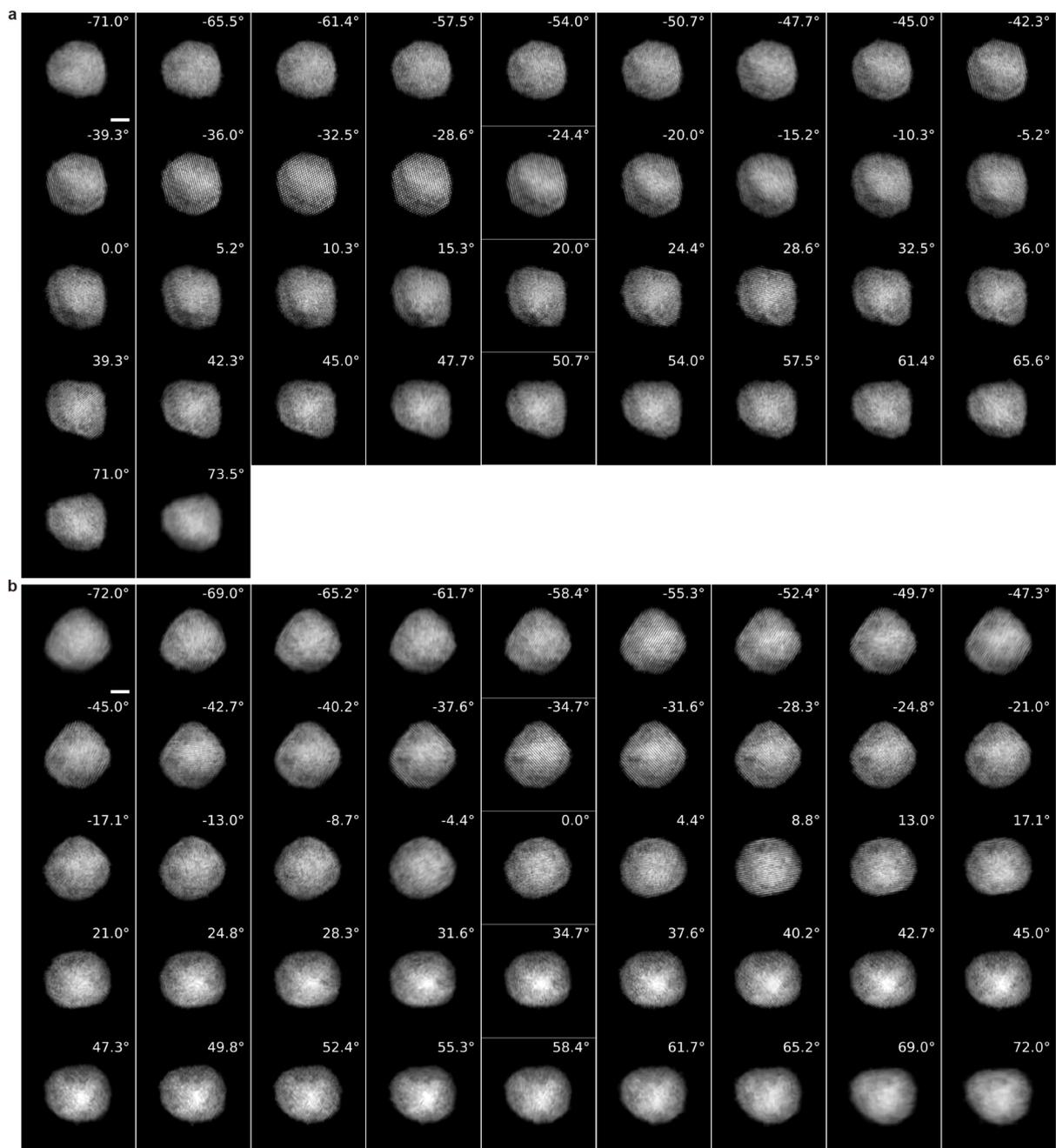

**Supplementary Figure 3 | Experimental tomographic tilt series of the 12k PtNi nanoparticles. a, b,** Post-processed ADF-STEM images at different tilt angles (white text in each figure), for PtNi-12k-p3 (**a**) and PtNi-12k-p4 (**b**). Scale bar, 2 nm.

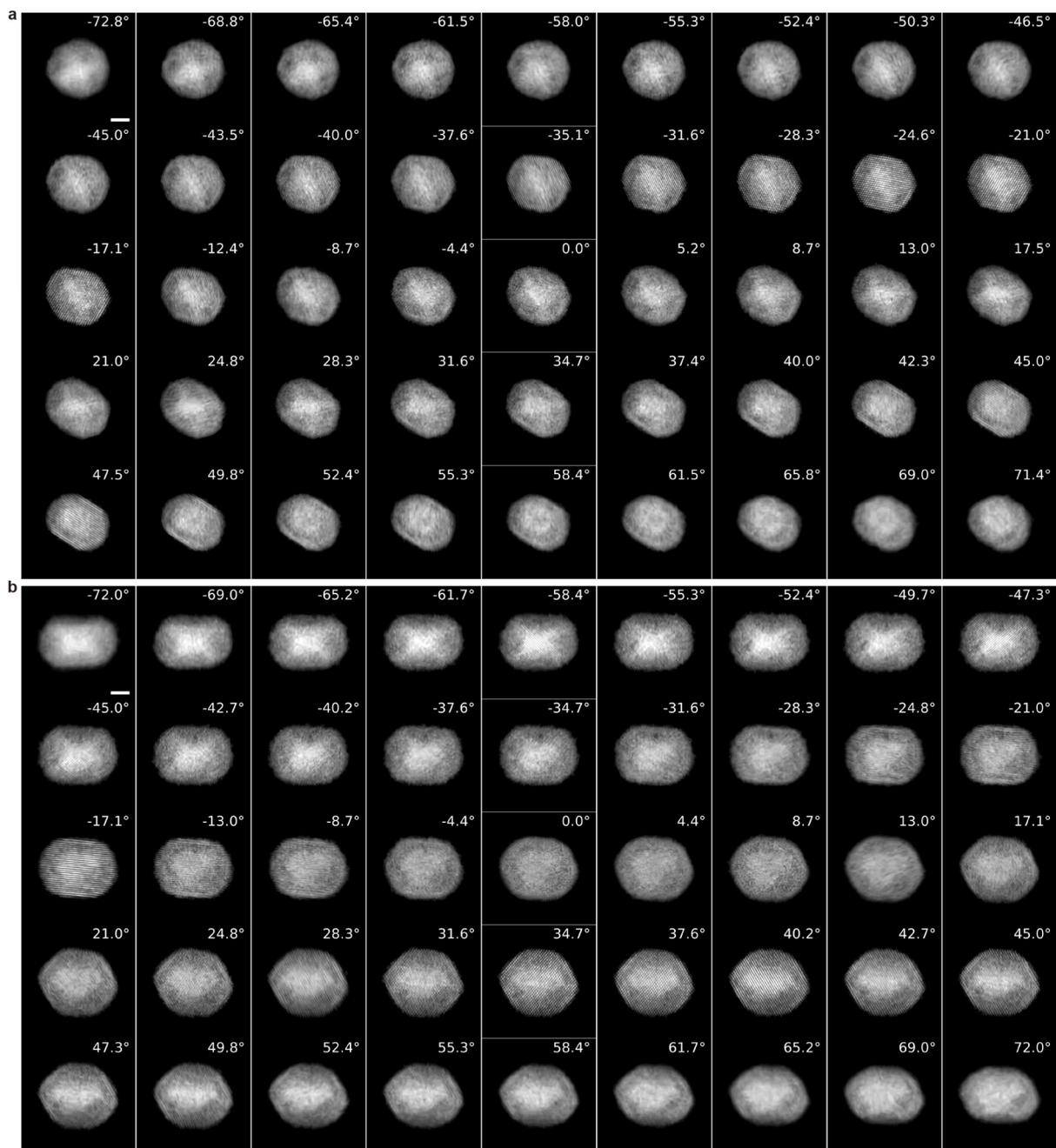

**Supplementary Figure 4 | Experimental tomographic tilt series of the 12k PtNi nanoparticles. a**, **b**, Post-processed ADF-STEM images at different tilt angles (white text in each figure), for PtNi-12k-p5 (**a**) and PtNi-12k-p6 (**b**). Scale bar, 2 nm.

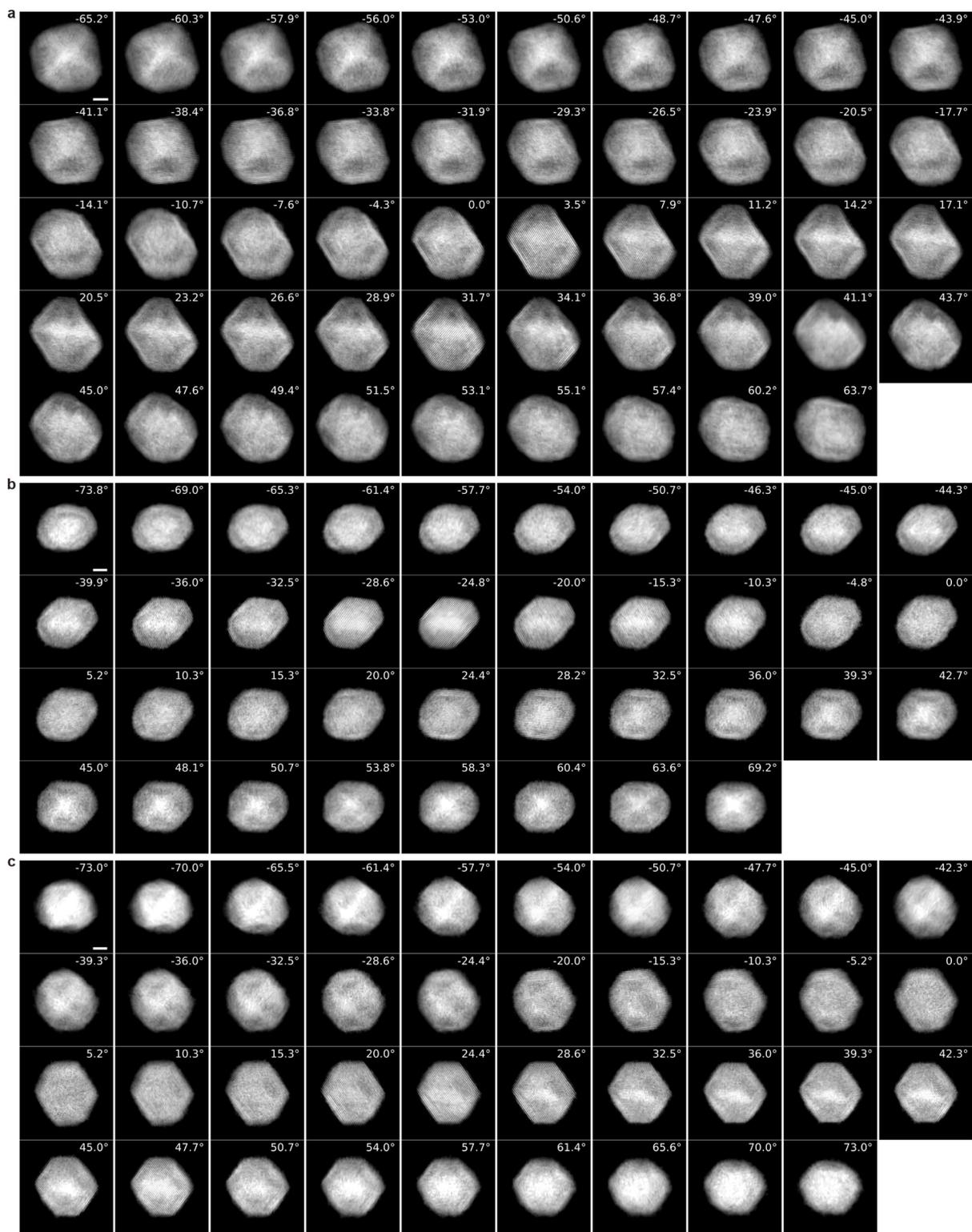

**Supplementary Figure 5 | Experimental tomographic tilt series of the pristine Ga-PtNi nanoparticles. a-c**, Post-processed ADF-STEM images at different tilt angles (white text in each figure), for Ga-PtNi-pristine-p1 (**a**), Ga-PtNi-pristine-p2 (**b**), and Ga-PtNi-pristine-p3 (**c**). Scale bar, 2 nm.

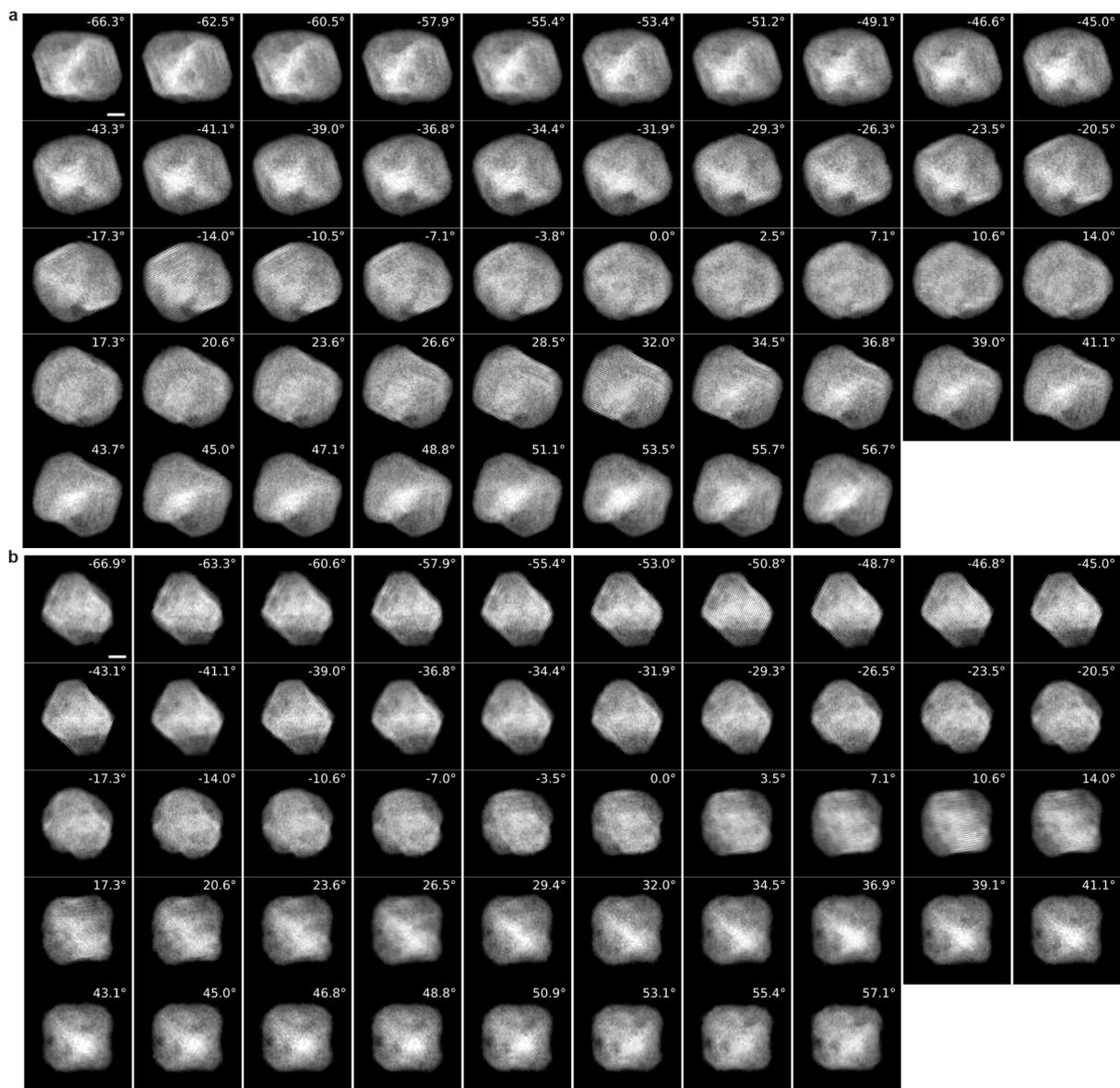

**Supplementary Figure 6 | Experimental tomographic tilt series of the 4k Ga-PtNi nanoparticles. a, b,** Post-processed ADF-STEM images at different tilt angles (white text in each figure), for Ga-PtNi-4k-p1 (**a**) and Ga-PtNi-4k-p2 (**b**). Scale bar, 2 nm.

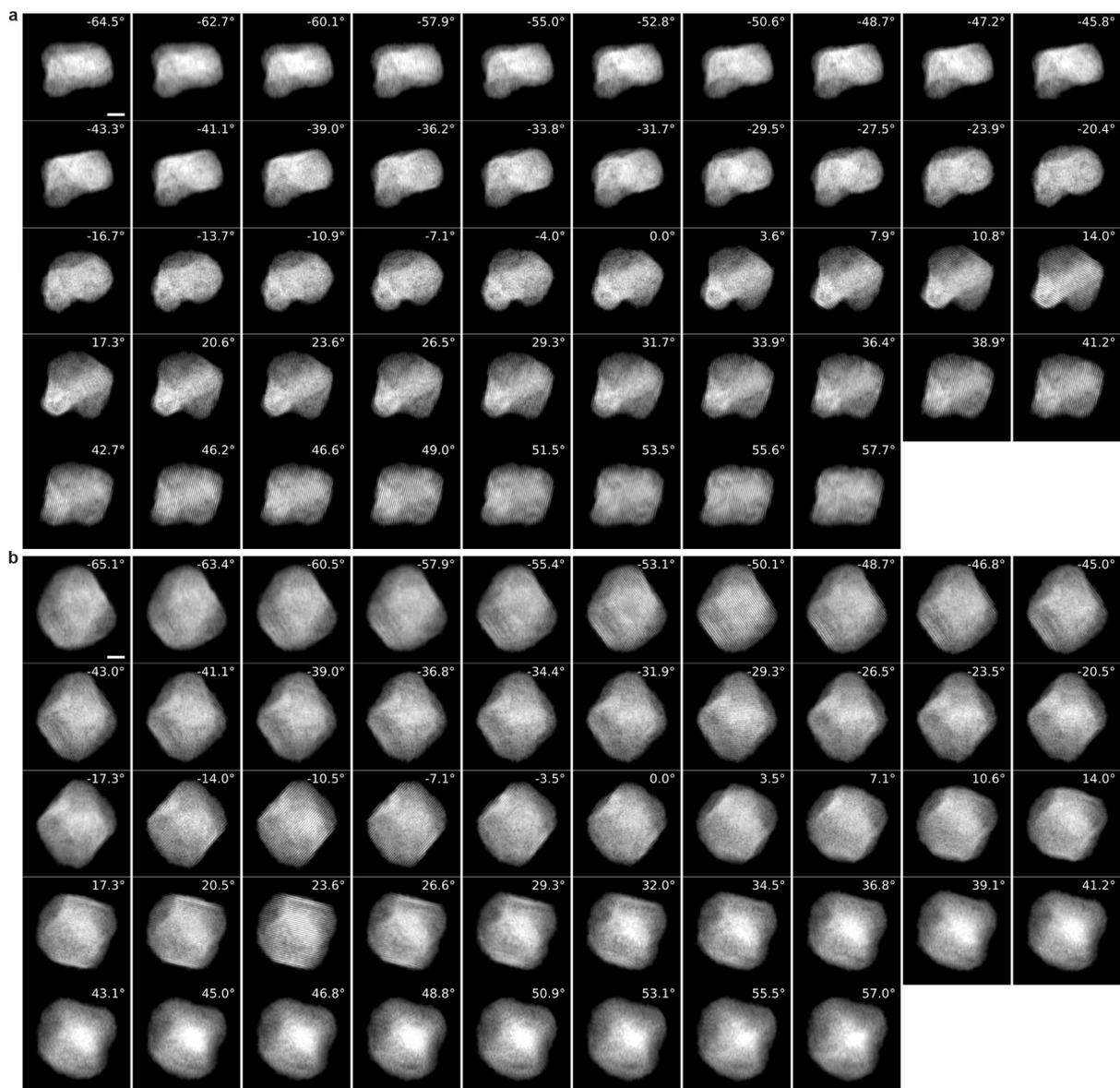

**Supplementary Figure 7 | Experimental tomographic tilt series of the 8k Ga-PtNi nanoparticles. a**, **b**, Post-processed ADF-STEM images at different tilt angles (white text in each figure), for Ga-PtNi-8k-p1 (**a**) and Ga-PtNi-8k-p2 (**b**). Scale bar, 2 nm.

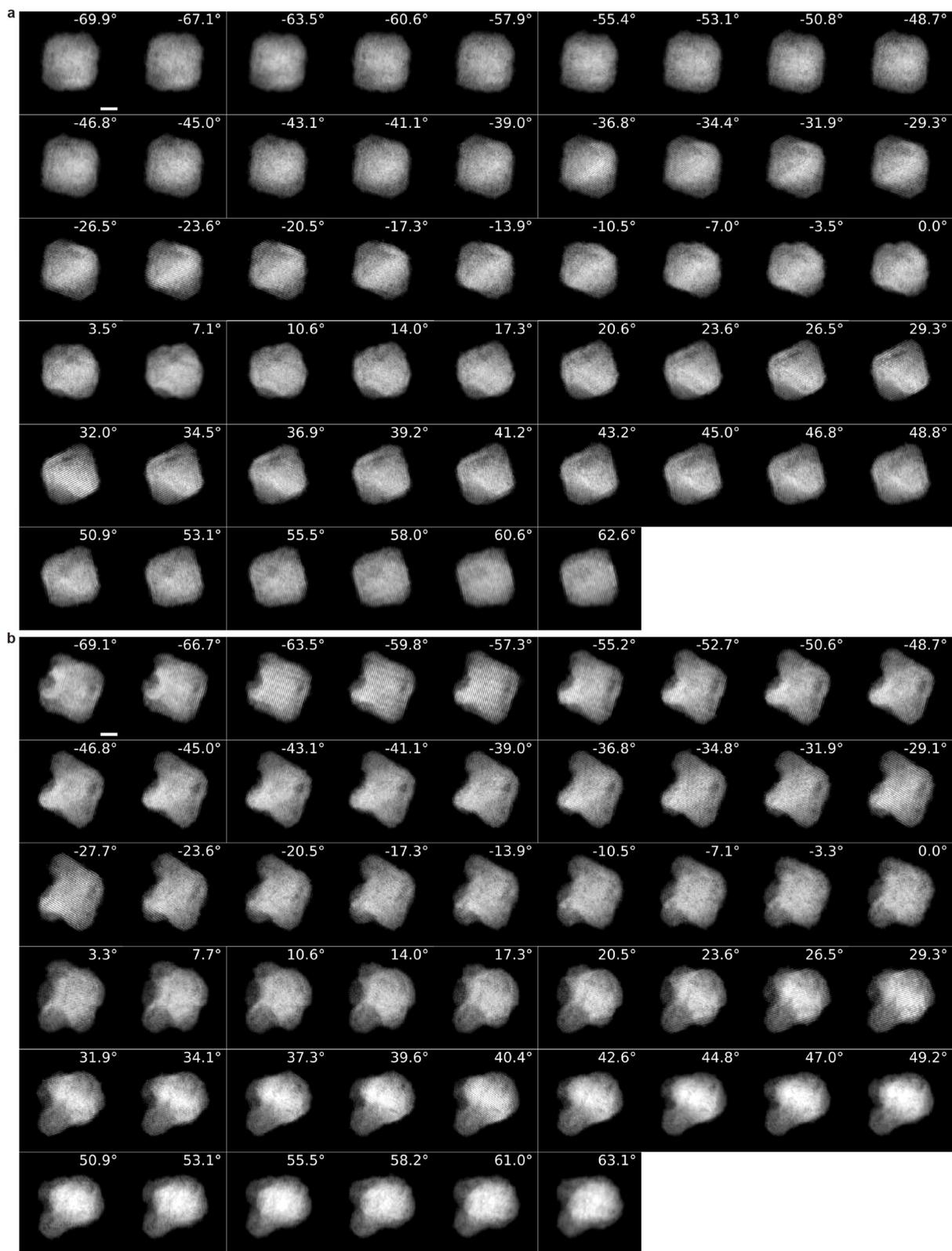

**Supplementary Figure 8 | Experimental tomographic tilt series of the 12k Ga-PtNi nanoparticles. a, b**, Post-processed ADF-STEM images at different tilt angles (white text in each figure), for Ga-PtNi-12k-p1 (**a**) and Ga-PtNi-12k-p2 (**b**). Scale bar, 2 nm.

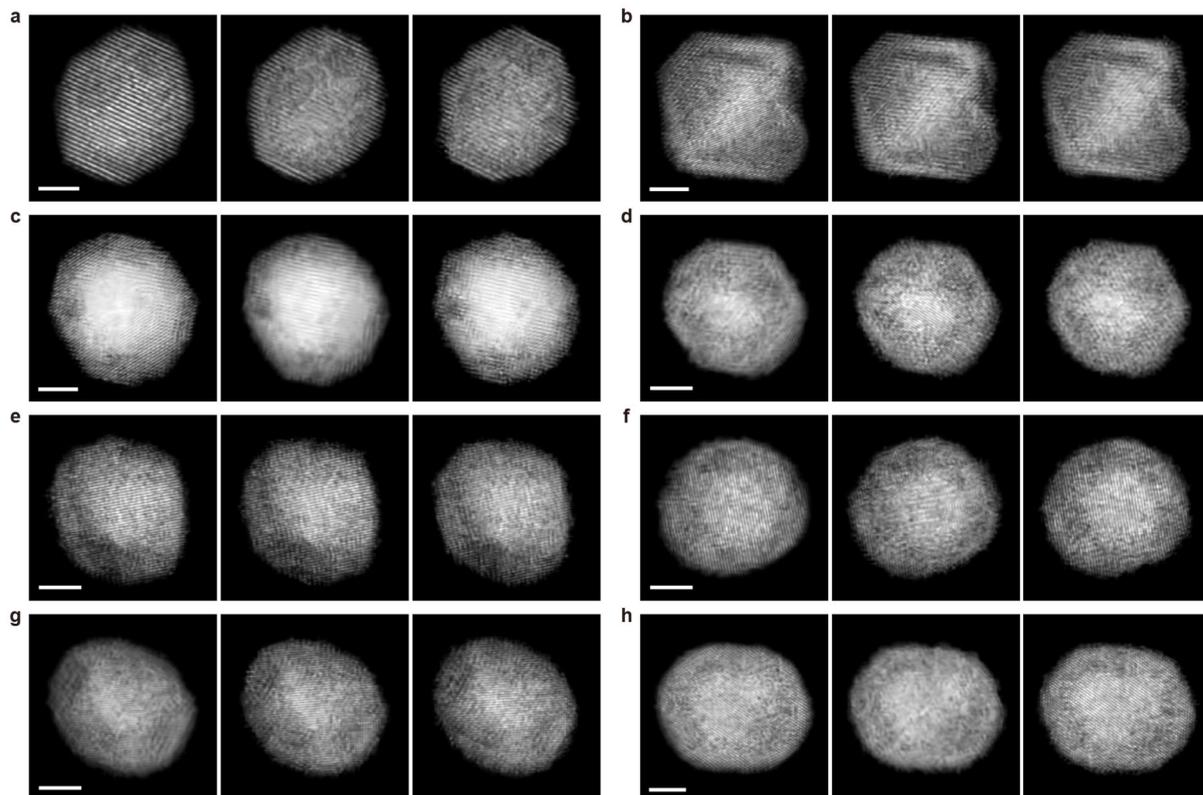

**Supplementary Figure 9 | Comparison of the zero-degree projections of PtNi nanoparticles during the AET experiments. a-h**, The ADF-STEM images at zero-degrees, obtained at the beginning, middle, and end of the experiment, for PtNi-pristine-p1 (**a**), PtNi-pristine-p2 (**b**), PtNi-12k-p1 (**c**), PtNi-12k-p2 (**d**), PtNi-12k-p3 (**e**), PtNi-12k-p4 (**f**), PtNi-12k-p5 (**g**), and PtNi-12k-p6 (**h**). Scale bar, 2nm.

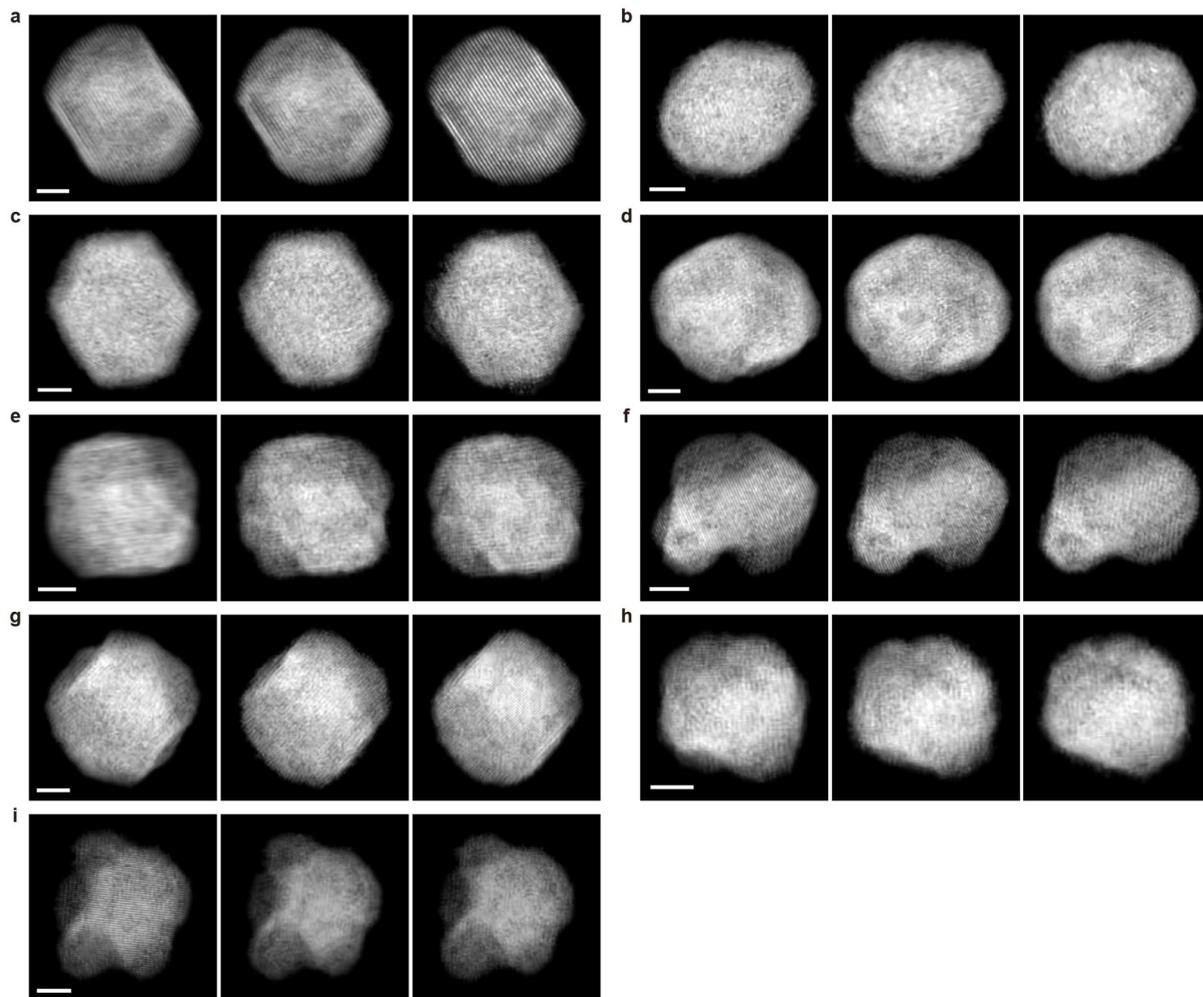

**Supplementary Figure 10 | Comparison of the zero-degree projections of Ga-PtNi nanoparticles during the AET experiments. a-i**, The ADF-STEM images at zero-degrees, obtained at the beginning, middle, and end of the experiment, for Ga-PtNi-pristine-p1 (**a**), Ga-PtNi-pristine-p2 (**b**), Ga-PtNi-pristine-p3 (**c**), Ga-PtNi-4k-p1 (**d**), Ga-PtNi-4k-p2 (**e**), Ga-PtNi-8k-p1 (**f**), Ga-PtNi-8k-p2 (**g**), Ga-PtNi-12k-p1 (**h**), and Ga-PtNi-12k-p2 (**i**). Scale bar, 2nm.

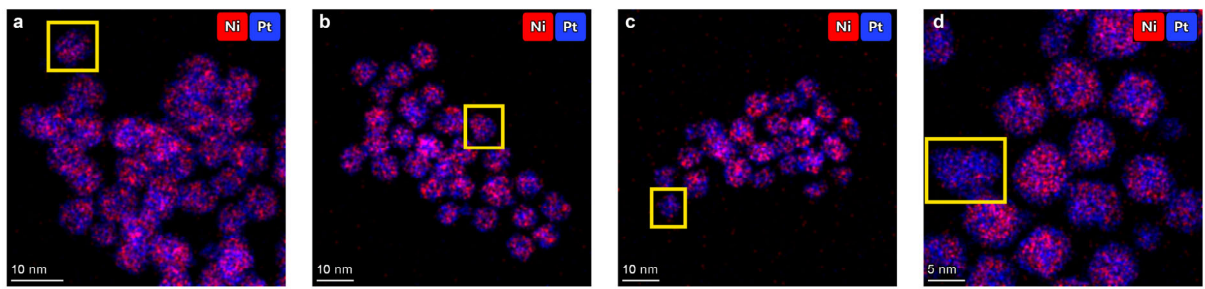

**Supplementary Figure 11 | Elemental mapping images of 12k cycled PtNi nanoparticles from EDS analysis.**
**a-d**, The background-corrected EDS elemental mapping images after Gaussian smoothing (sigma parameter = 2.0 in Velox 3.12.1.5, an FEI TEM imaging software). Individual nanoparticles analyzed for composition are indicated by yellow boxes. The nanoparticles highlighted by yellow boxes exhibit Pt atomic percentages of 65 at% (**a**), 64 at% (**b**), 77 at% (**c**), and 80 at% (**d**).

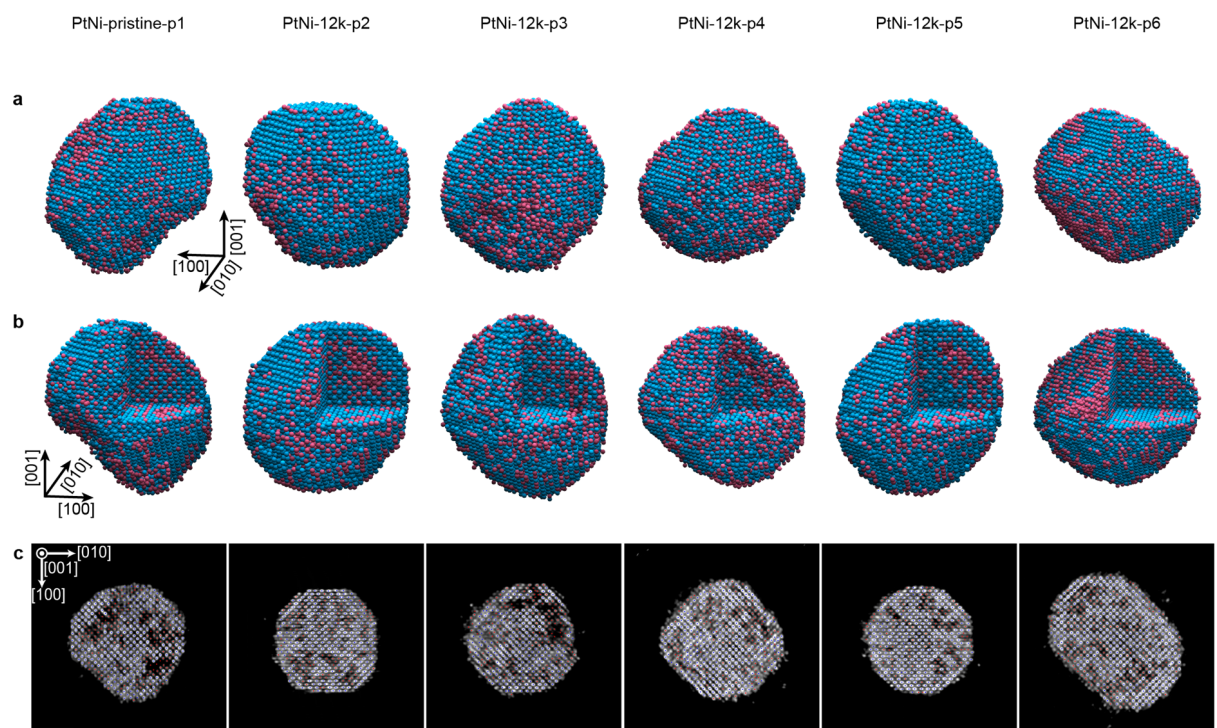

**Supplementary Figure 12 | Experimentally determined 3D atomic structures of PtNi nanoparticles during potential cycles. a**, Overall 3D atomic structure of PtNi (pristine and 12k) after given number of potential cycles. Note that these are the measured PtNi nanoparticles (PtNi-pristine-p1, PtNi-12k-p2, PtNi-12k-p3, PtNi-12k-p4, PtNi-12k-p5, PtNi-12k-p6) not illustrated in Fig. 1a-c. **b**, 3D atomic structures after a 180-degree rotation along the [001] axis from those in (**a**), with one octant of the nanoparticles removed to reveal the internal atomic structure. **c**, 1 Å thick internal slices at the center of the 3D tomograms perpendicular to the [001] direction. The intensity is shown in grayscale, with the atomic coordinates of Pt and Ni marked by blue and red dots, respectively. Scale bar, 2 nm.

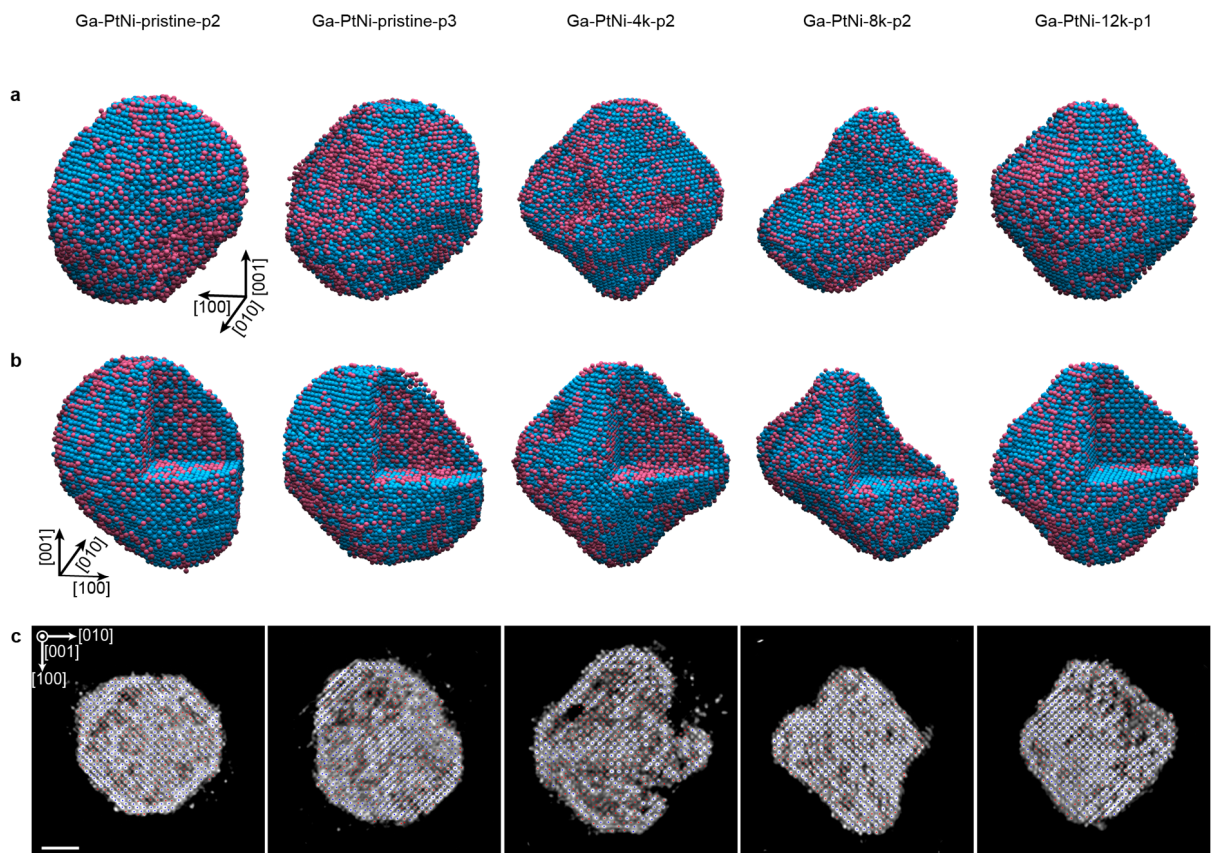

**Supplementary Figure 13 | Experimentally determined 3D atomic structures of Ga-PtNi nanoparticles during potential cycles. a**, Overall 3D atomic structure of Ga-PtNi (pristine, 4k, 8k, and 12k) after given number of potential cycles. Note that these are the remaining Ga-PtNi nanoparticles (Ga-PtNi-pristine-p2, Ga-PtNi-pristine-p3, Ga-PtNi-4k-p2, Ga-PtNi-8k-p2, Ga-PtNi-12k-p1), excluding the representative nanoparticles specified in Fig. 1a-c. **b**, 3D atomic structures after a 180-degree rotation along the [001] axis from those in (**a**), with one octant of the nanoparticles removed to reveal the internal atomic structure. **c**, 1 Å thick internal slices at the center of the 3D tomograms perpendicular to the [001] direction. The intensity is shown in grayscale, with the atomic coordinates of Pt and Ni marked by blue and red dots, respectively. Scale bar, 2 nm.

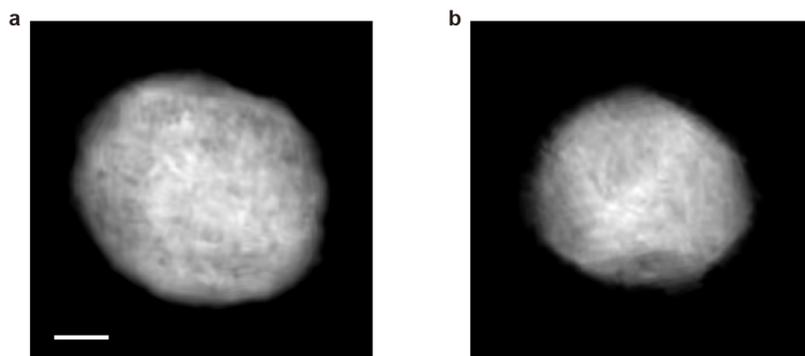

**Supplementary Figure 14 | Example experimental 2D projections where 3D octahedral-shaped nanoparticles appear spherical. a**, **b**, Experimental projection images of Ga-PtNi-pristine-p1 at 60.2° (**a**) and Ga-PtNi-pristine-p2 at −65.5° (**b**), where the distinct octahedral shape evident in 3D tomogram is not clearly visible and appears spherical.

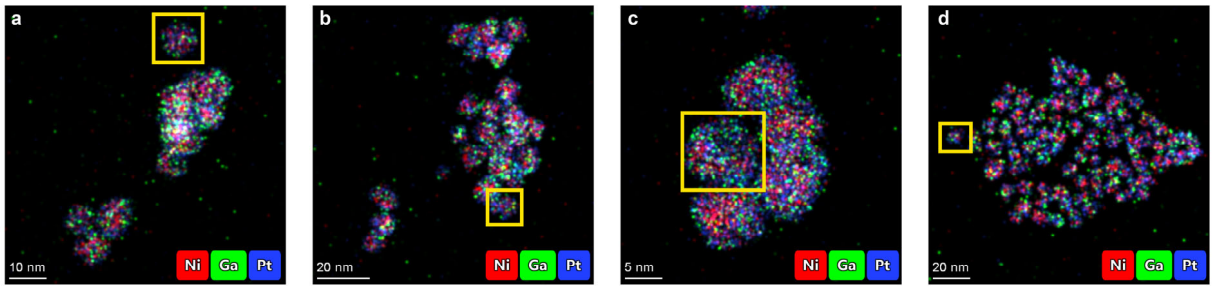

**Supplementary Figure 15 | Elemental mapping images of 12k cycled Ga-PtNi nanoparticles from EDS analysis. a-d**, The background-corrected EDS elemental mapping images after Gaussian smoothing (sigma parameter = 2.0 in Velox 3.12.1.5, an FEI TEM imaging software). Individual nanoparticles analyzed for composition are indicated by yellow boxes. The selected nanoparticles exhibit Pt atomic percentages of 65 at% (**a**), 64 at% (**b**), 64 at% (**c**), and 69 at% (**d**), respectively. These nanoparticles represent the highest Pt atomic percentages observed within each image in (**a-d**).

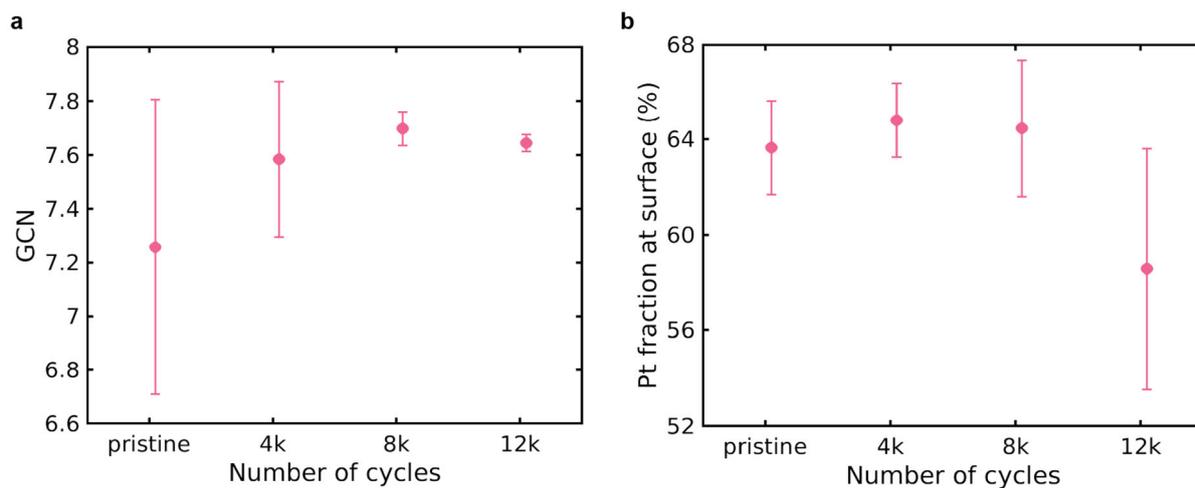

**Supplementary Figure 16 | Quantitative characterization of GCN and chemical compositions of the Ga-PtNi nanoparticles during potential cycles. a**, **b**, Average of GCN (**a**), and Pt fraction at surface (**b**) during potential cycles of Ga-PtNi nanoparticles. Note that the values and error bars shown in (**a**, **b**) correspond to the averages and standard deviations calculated from multiple samples: 3 Ga-PtNi pristine, 2 Ga-PtNi 4k, 2 Ga-PtNi 8k, and 2 Ga-PtNi 12k nanoparticles. Source data for dot or line plots in this figure are provided as a Source Data file.

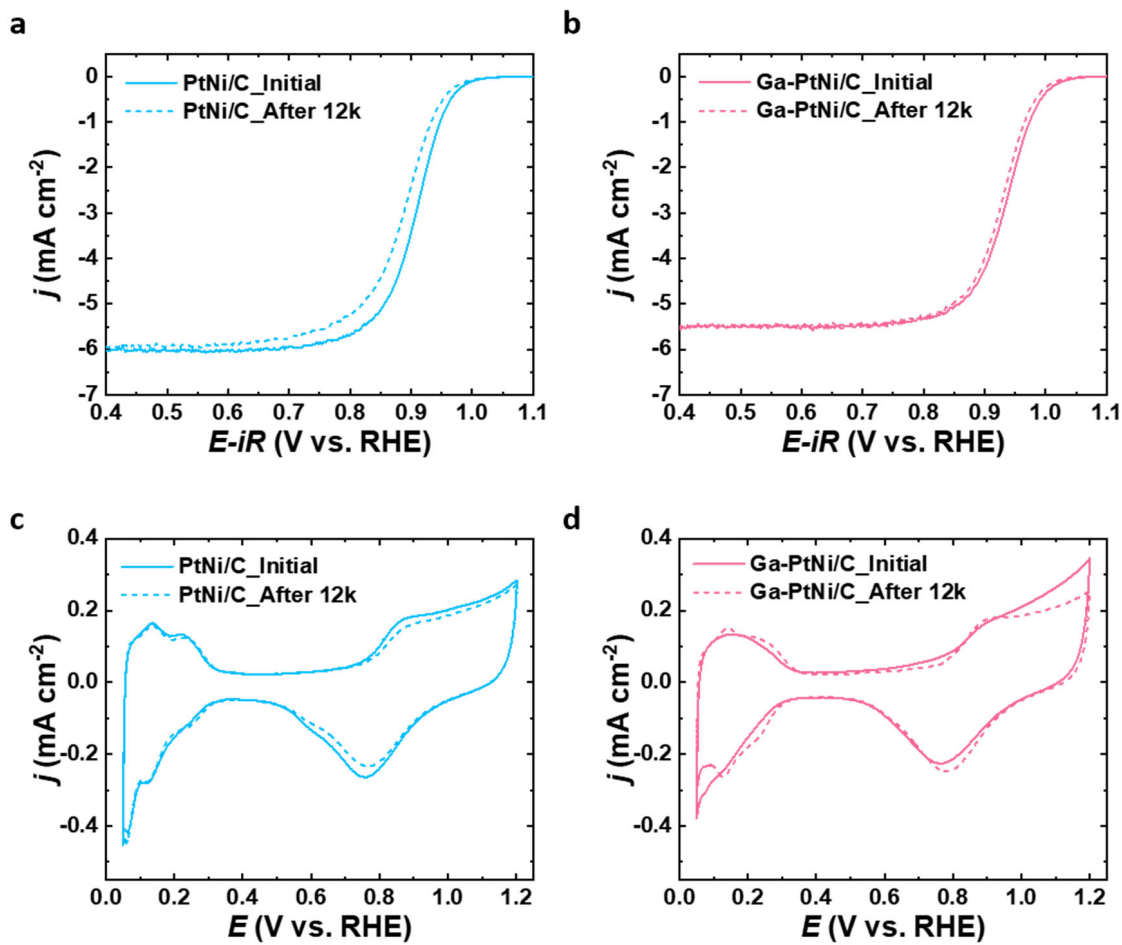

**Supplementary Figure 17 | Electrochemical characterization of PtNi/C and Ga-PtNi/C. a**, **b**, ORR polarization curves of (**a**) PtNi/C and (**b**) Ga-PtNi/C. **c**, **d**, Cyclic voltammograms of (**c**) PtNi/C and (**d**) Ga-PtNi/C. The catalyst electrodes were coated onto a glassy carbon electrode (GCE, 0.1963 cm$^2$) with a Pt loading of 10 μ$g_{Pt}$ cm$_{geo}^{-2}$. ORR polarization curves were recorded in an oxygen-saturated 0.1 M HClO$_4$ solution over the potential range from 0.05 to 1.1 V vs. RHE with a scan rate of 20 mV s$^{-1}$ at a rotating speed of 1600 rpm. Cyclic voltammetry was performed in an argon-saturated 0.1 M HClO$_4$ solution between 0.05 and 1.2 V vs. RHE with a scan rate of 50 mV s$^{-1}$. The resistance (R, 24.0 ± 0.5 Ω) of the electrochemical cell was measured by electrochemical impedance spectroscopy. The abbreviations of *E* (*x*-axis) and *j* (*y*-axis) represent the potential and the current density, respectively. The potentials of the ORR polarization curves are iR-corrected (*E-iR*), and those of the cyclic voltammograms not iR-corrected (*E*). Source data for dot or line plots in this figure are provided as a Source Data file.

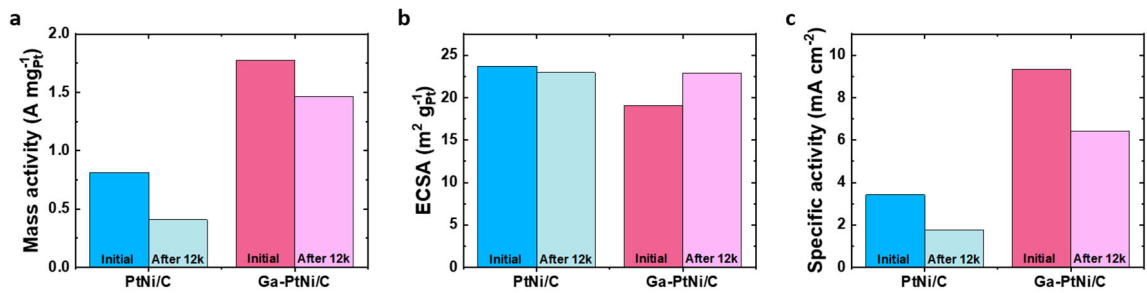

**Supplementary Figure 18 | Electrochemical performance of PtNi/C and Ga-PtNi/C. a-c**, Mass activities (**a**), ECSAs (**b**), and specific activities (**c**) of PtNi/C and Ga-PtNi/C before and after 12k durability cycles.

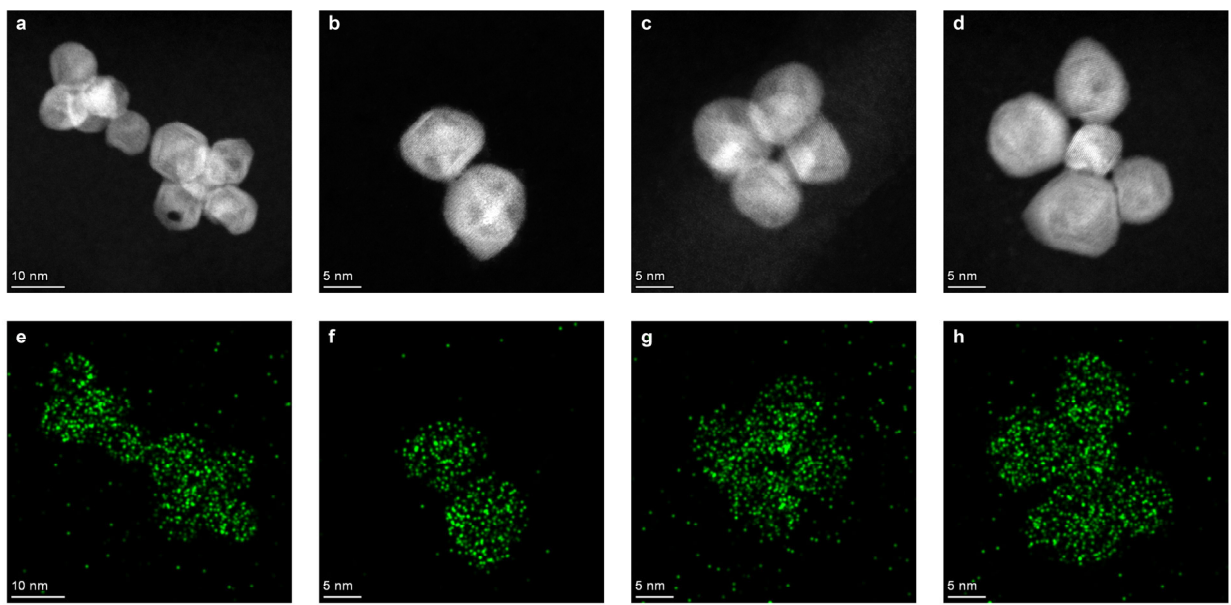

**Supplementary Figure 19 | ADF-STEM images and Ga mapping images of Ga-PtNi nanoparticles in the pristine state. a-d,** ADF-STEM images of Ga-PtNi nanoparticles in the pristine state. **e-h,** Background-corrected EDS elemental mapping images for Ga after Gaussian smoothing (sigma parameter = 2.0 in Velox 3.12.1.5, an FEI TEM imaging software). Each image in (**e-h**) corresponds to the images in the upper row (**a-d**).

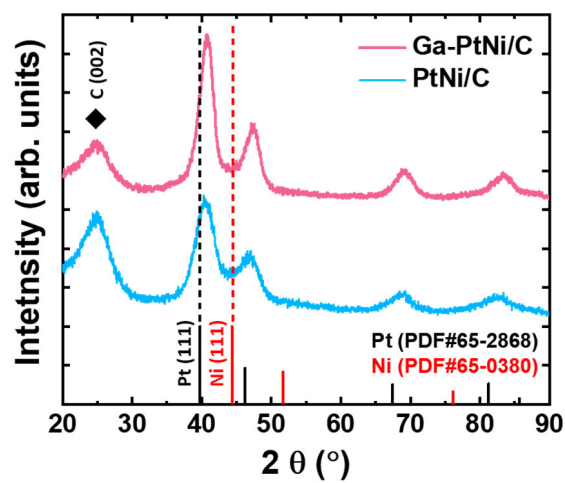

**Supplementary Figure 20 | XRD patterns of PtNi/C and Ga-PtNi/C.** The black diamond, black dashed line, and red dashed line indicate the expected diffraction peak positions for graphite carbon (002), Pt (111), and Ni (111), respectively. To confirm the presence of alloyed Pt phases without Ni or Ga segregation, the expected diffraction peak positions of Pt (PDF#65-2868) and Ni (PDF#65-0380) are overlaid with black and red lines, respectively. Source data for dot or line plots in this figure are provided as a Source Data file.

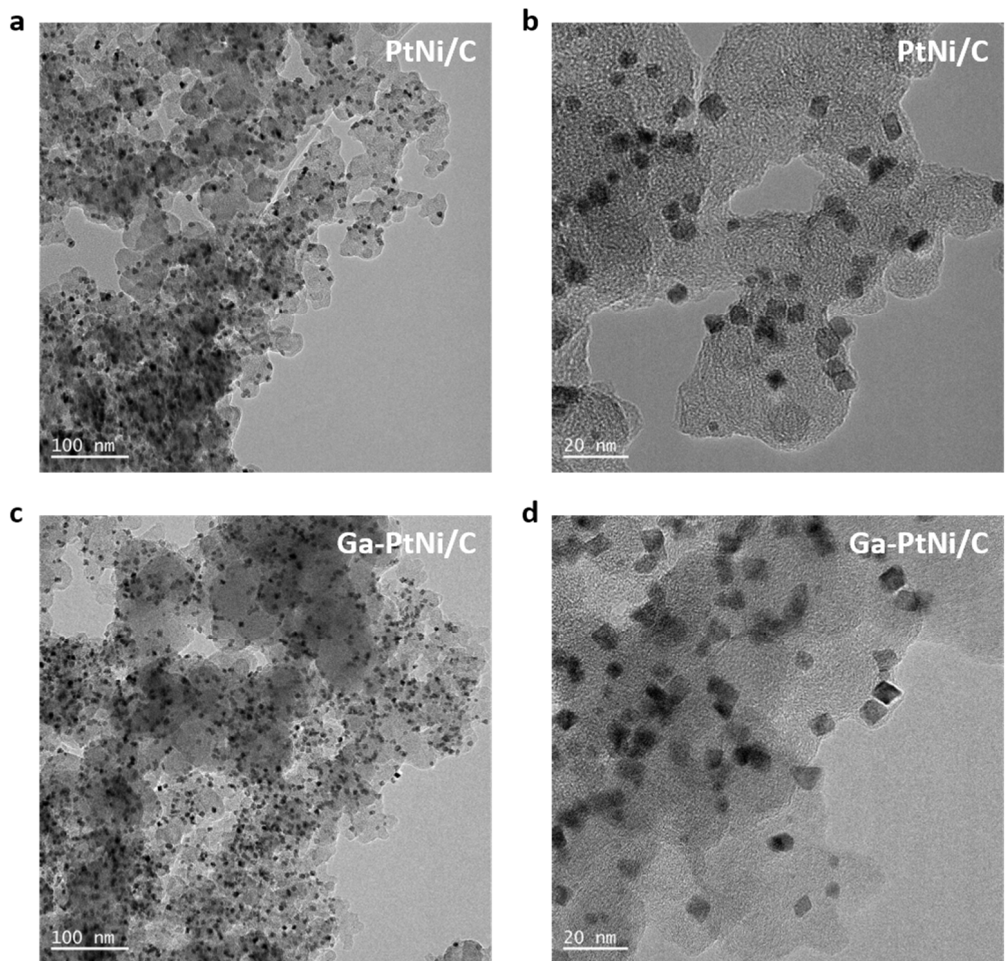

**Supplementary Figure 21 | TEM images of PtNi/C and Ga–PtNi/C. a-d**, TEM images of PtNi/C (**a, b**) and Ga-PtNi/C (**c, d**) at two different magnifications.

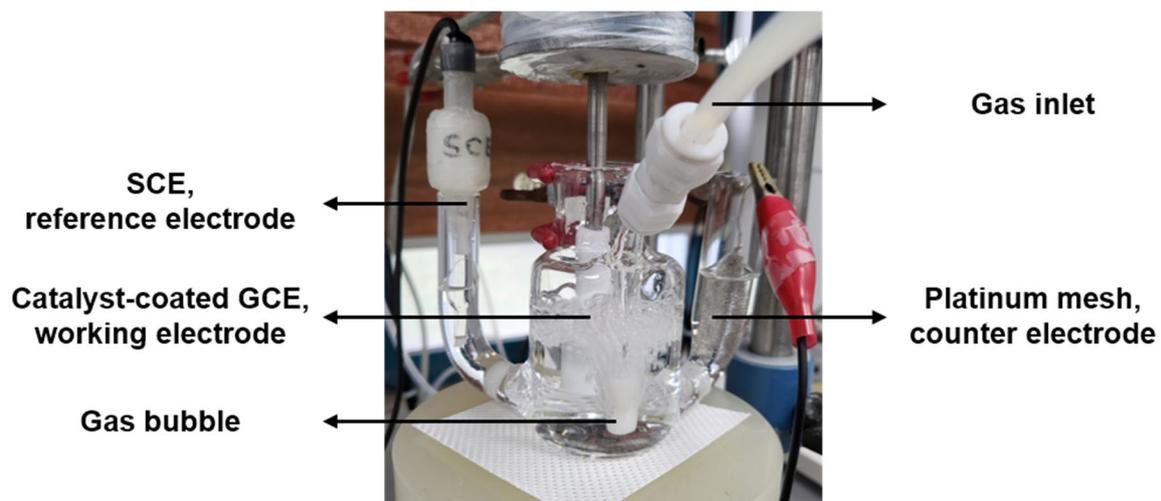

**Supplementary Figure 22 | The electrochemical evaluation setup with a three-electrode system.**

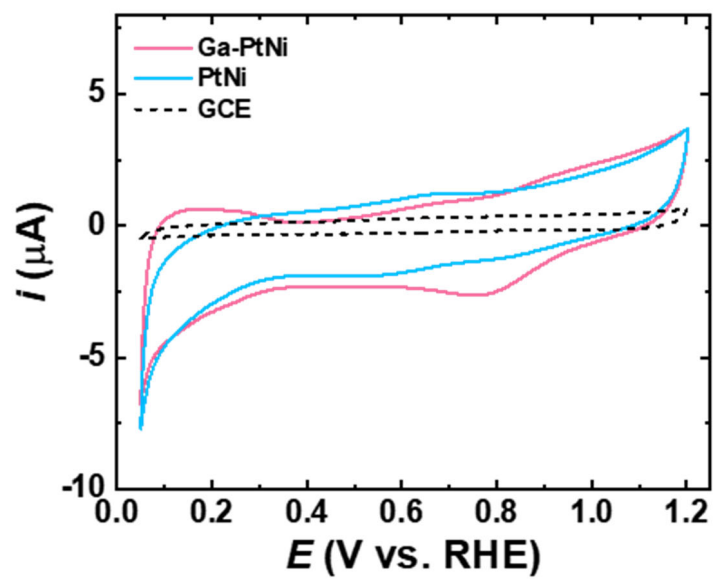

**Supplementary Figure 23 | Cyclic voltammograms of PtNi and Ga-PtNi nanoparticles.** Cyclic voltammograms for the PtNi- and Ga-PtNi-coated TEM grids on GCE and bare GCE, recorded in an Ar-saturated 0.1 M HClO$_4$ solution over a potential range from 0.05 to 1.2 V vs. RHE at a scan rate of 50 mV s$^{-1}$. The abbreviations of $E$ ($x$-axis) and $i$ ($y$-axis) represent the potential and the current, respectively. Source data for dot or line plots in this figure are provided as a Source Data file.

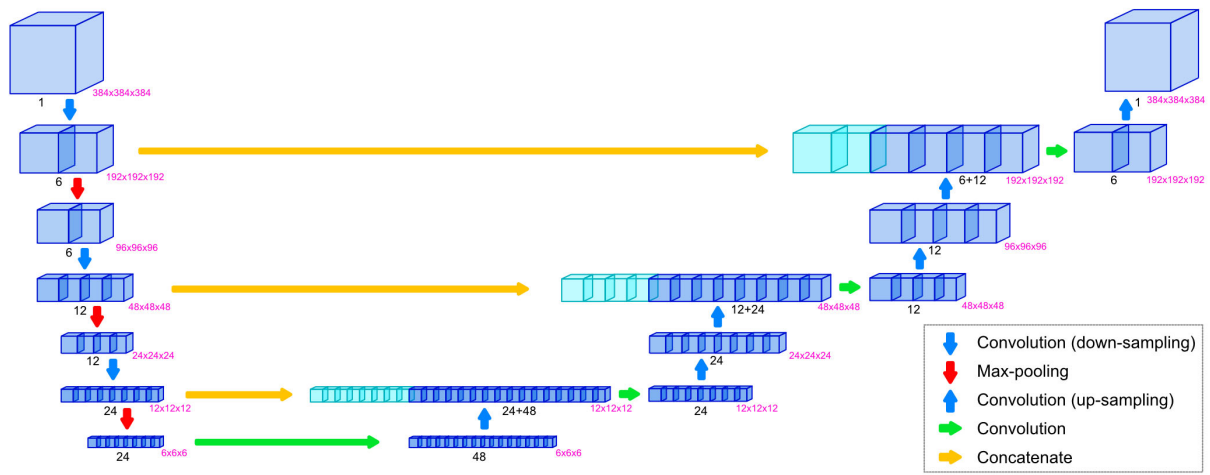

**Supplementary Figure 24 | Architecture of the deep learning augmentation neural network.** The neural network framework is based on a 3D U-Net[1]. The boxes represent 3D feature maps. The black numbers below each feature map indicate the number of channels, and the pink numbers next to them denote the volume size of the 3D feature maps.

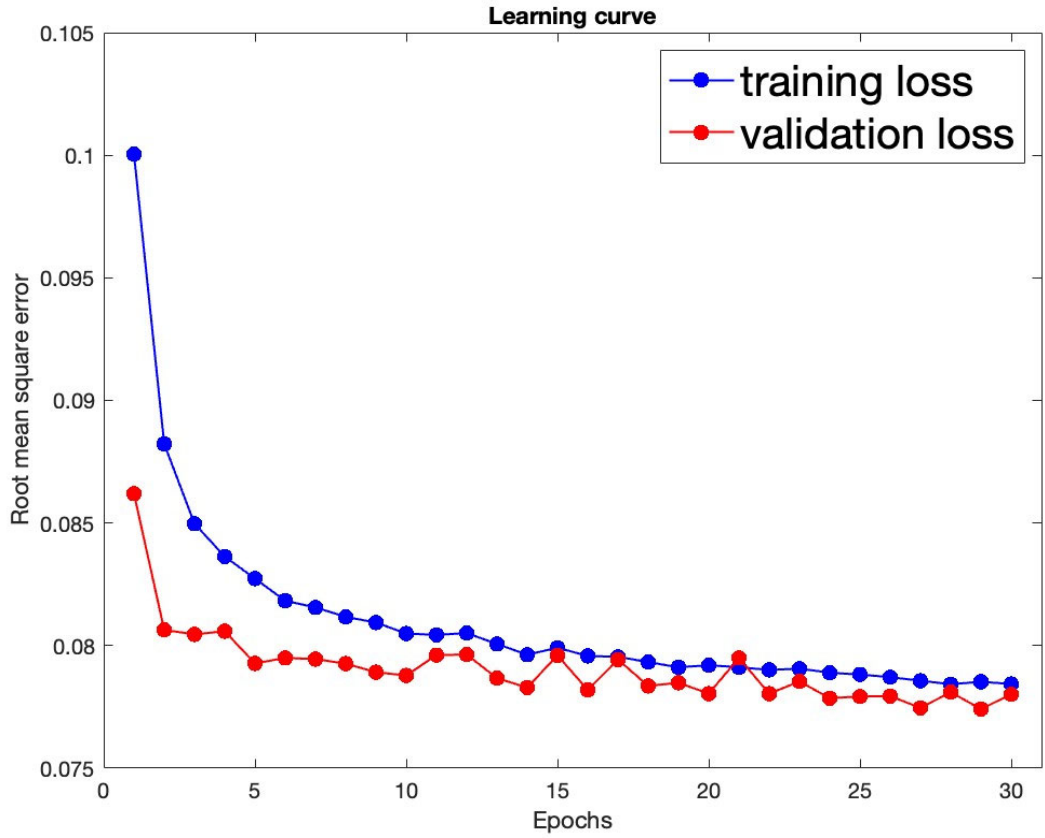

**Supplementary Figure 25 | Learning curves during the training of the deep learning neural network.** The blue dots and lines represent the root mean square error losses for the training dataset, and red dots and lines represent the losses for the validation dataset. Source data for dot or line plots in this figure are provided as a Source Data file.

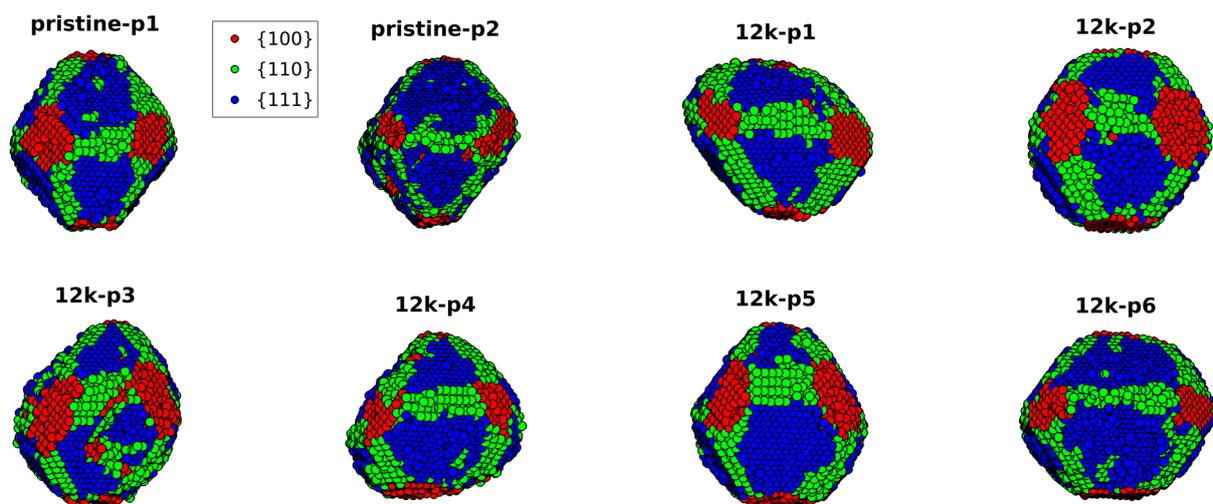

**Supplementary Figure 26 | Distribution of surface facets of the PtNi nanoparticles during potential cycles.** Distribution of surface atoms for two particles of the pristine state and six particles after 12k cycles, which are classified into the three dominant facet families of {100}, {110} and {111}. Red, green and blue dots represent the atom positions that are assigned to {100}, {110} and {111}, respectively.

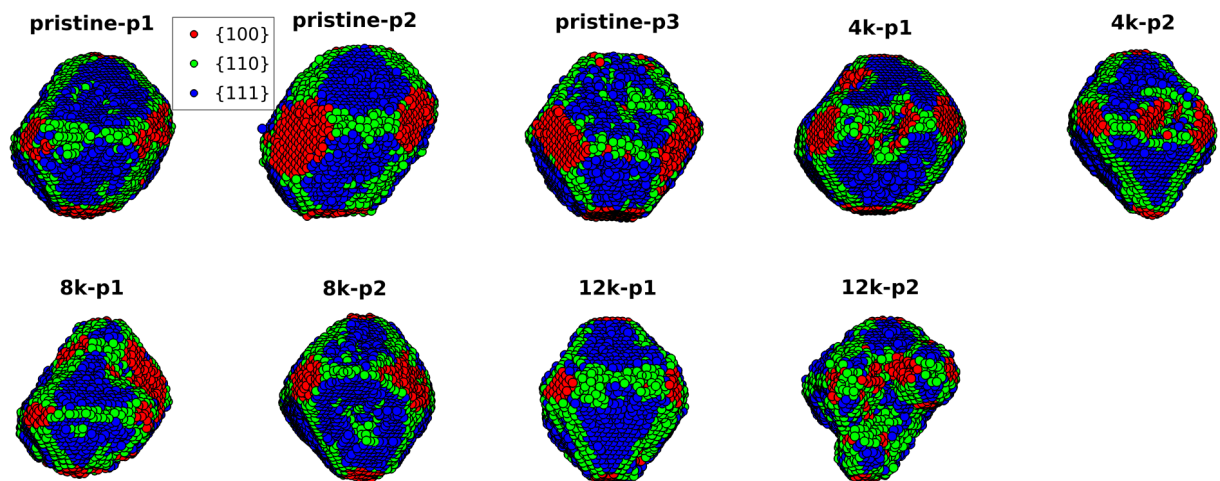

**Supplementary Figure 27 | Distribution of surface facets of the Ga-PtNi nanoparticles during potential cycles.** Distribution of surface atoms for the pristine state (three particles) and after 4k, 8k, and 12k cycles (two particles for each cycle), which are classified into the three dominant facet families of {100}, {110} and {111}. Red, green and blue dots represent the atom positions that are assigned to {100}, {110} and {111}, respectively.

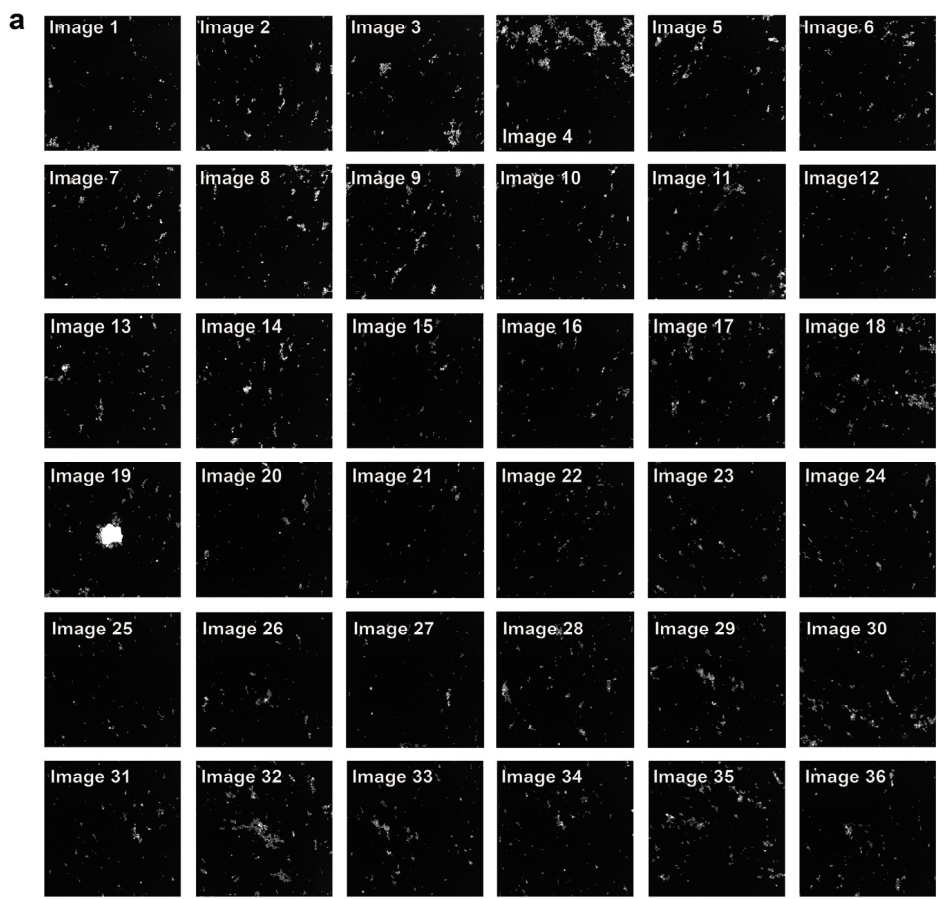

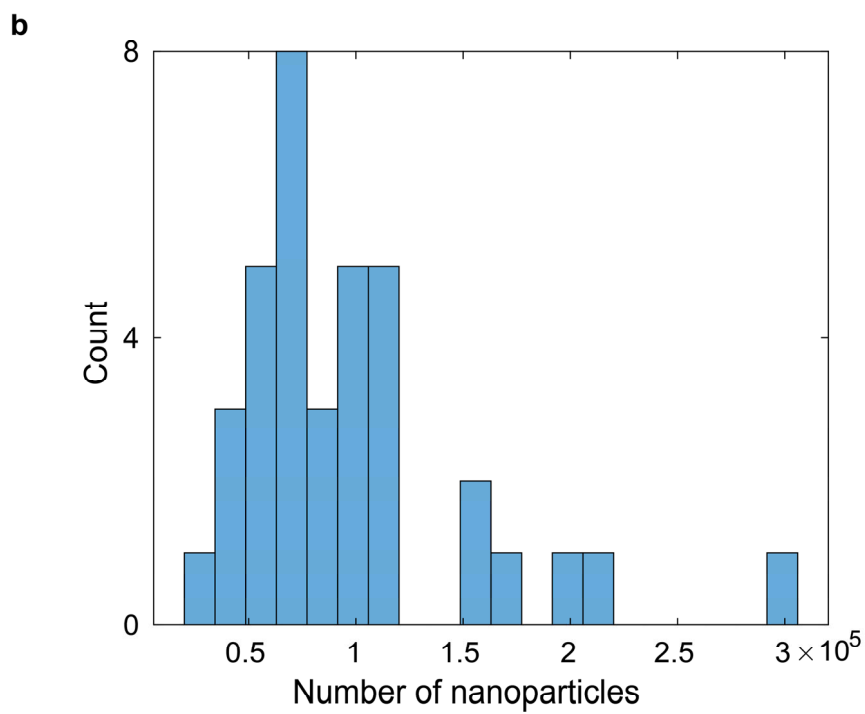

**Supplementary Figure 28 | Low-magnification ADF-STEM images and histogram showing the number of nanoparticles per image for PtNi nanoparticles. a,** 36 low-magnification ADF-STEM images of the PtNi sample, each with a field of view of 19.0 × 19.0 μm$^2$. **b,** Histogram representing the number of nanoparticles per image for the images in (**a**).

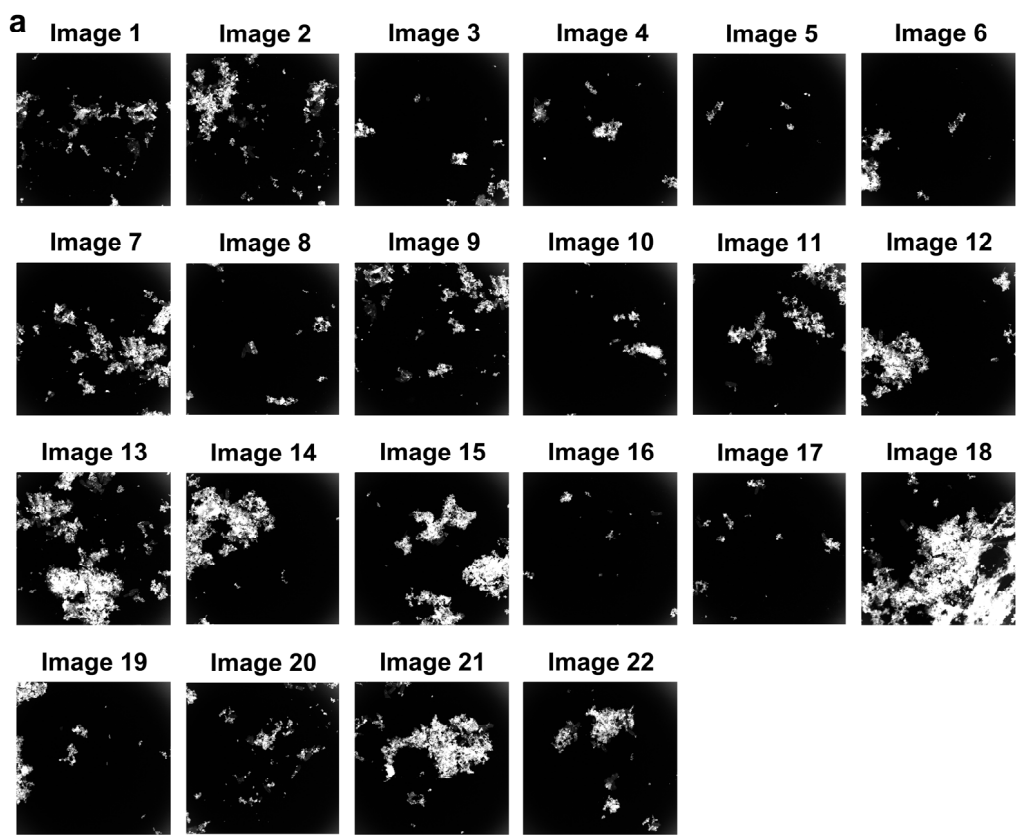

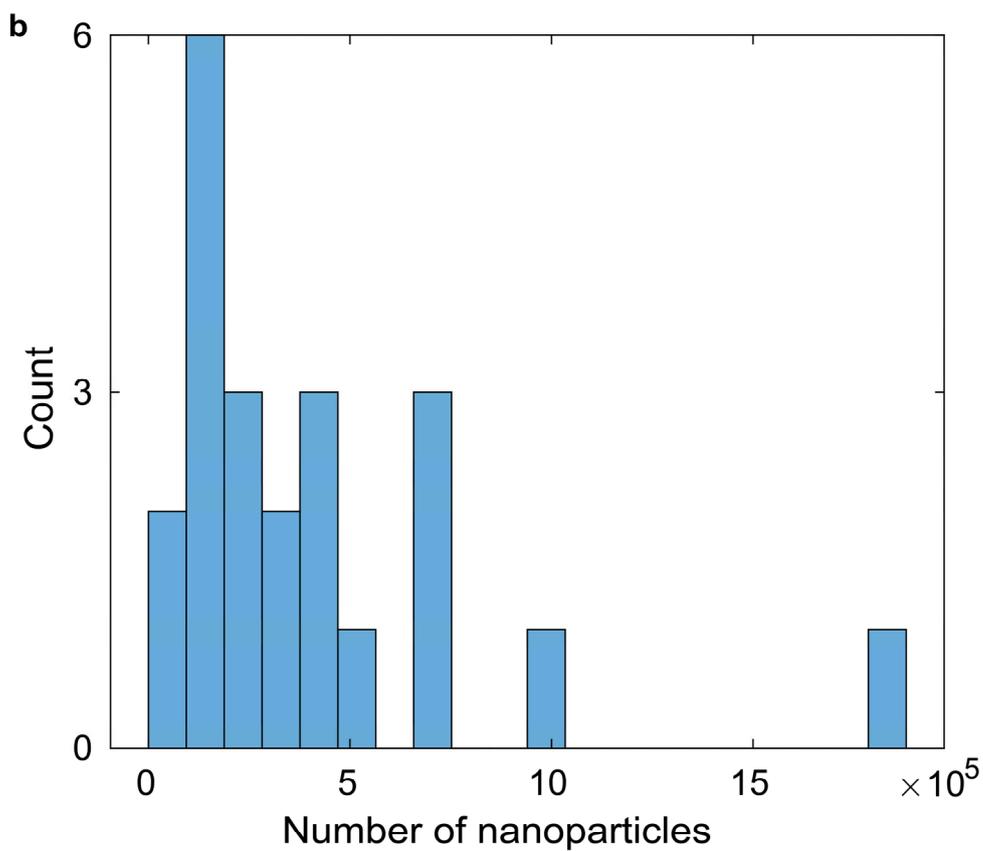

**Supplementary Figure 29 | Low-magnification ADF-STEM images and histogram showing the number of nanoparticles per image for Ga-PtNi nanoparticles. a,** 22 low-magnification ADF-STEM images of the Ga-PtNi sample, each with a field of view of 13.4 × 13.4 μm². **b,** Histogram representing the number of nanoparticles per image for the images in (**a**).

**Supplementary Table 1 | Experimental conditions, analysis parameters and tracing results of PtNi nanoparticles.**

|  | PtNi pristine | | PtNi 12k | | | | | |
|---|---|---|---|---|---|---|---|---|
|  | PtNi-pristine-p1 | PtNi-pristine-p2[a] | PtNi-12k-p1[a] | PtNi-12k-p2 | PtNi-12k-p3 | PtNi-12k-p4 | PtNi-12k-p5 | PtNi-12k-p6 |
| **STEM data acquisition** | | | | | | | | |
| Electron microscope type | FEI Spectra Ultra at KARA | FEI Spectra Ultra at KARA | FEI Spectra Ultra at KARA | FEI Spectra Ultra at KARA | FEI Spectra Ultra at KARA | FEI Spectra Ultra at KARA | FEI Spectra Ultra at KARA | FEI Spectra Ultra at KARA |
| Acceleration voltage (kV) | 300 | 300 | 300 | 300 | 300 | 300 | 300 | 300 |
| Convergence semi-angle (mrad) | 21.4 | 21.4 | 21.4 | 21.4 | 21.4 | 21.4 | 21.4 | 21.4 |
| Detector inner angle (mrad) | 39 | 39 | 39 | 39 | 39 | 39 | 39 | 39 |
| Detector outer angle (mrad) | 200 | 200 | 200 | 200 | 200 | 200 | 200 | 200 |
| Pixel size (Å) | 0.331 | 0.331 | 0.328 | 0.328 | 0.328 | 0.328 | 0.328 | 0.328 |
| Screen current (pA) | 10 | 10 | 10 | 10 | 20 | 10 | 10 | 10 |
| Dwell time (µs) | 3 | 3 | 3 | 3 | 3 | 3 | 3 | 3 |
| # of consecutive images | 3 | 3 | 3 | 3 | 3 | 3 | 3 | 3 |
| # of projections | 41 | 41 | 41 | 42 | 38 | 45 | 45 | 45 |
| Tilt angle range (°) | −71.0 +71.0 | −71.0 +71.0 | −71.0 +71.0 | −71.0 +74.1 | −71.0 +73.5 | −72.0 +72.0 | −72.8 +71.4 | −72.0 +72.0 |
| Electron dose ($10^5$ $e$/Å$^2$) | 2.1 | 2.1 | 2.1 | 2.2 | 4.0 | 2.4 | 2.4 | 2.4 |
| **3D reconstruction** | | | | | | | | |
| Algorithm | GENFIRE | GENFIRE | GENFIRE | GENFIRE | GENFIRE | GENFIRE | GENFIRE | GENFIRE |
| Interpolation method | DFT | DFT | DFT | DFT | DFT | DFT | DFT | DFT |
| # of iterations | 1000 | 1000 | 1000 | 1000 | 1000 | 1000 | 1000 | 1000 |
| Interpolation radius (pixel) | 0.1 | 0.1 | 0.1 | 0.1 | 0.1 | 0.1 | 0.1 | 0.1 |
| Oversampling ratio | 4 | 4 | 4 | 4 | 4 | 4 | 4 | 4 |
| **Tracing result** | | | | | | | | |
| # of atoms | | | | | | | | |
| Ni | 3954 | 7102 | 2623 | 3048 | 3510 | 3926 | 3270 | 5148 |
| Pt | 6792 | 10404 | 9486 | 6214 | 6189 | 6343 | 6171 | 8482 |
| B factor (Å$^2$) | | | | | | | | |
| Ni | 6.4 | 5.7 | 7.5 | 6.7 | 5.5 | 6.6 | 6.4 | 5.8 |
| Pt | 7.0 | 7.6 | 8.0 | 7.3 | 6.6 | 7.9 | 7.8 | 7.2 |
| R factor (%) | 10.8 | 10.1 | 10.5 | 10.0 | 11.3 | 10.4 | 9.8 | 10.0 |
| **Composition** | | | | | | | | |
| Pt atomic percent (at%) | 63 | 59 | 78 | 67 | 64 | 62 | 65 | 62 |
| **Precision estimation** | | | | | | | | |
| Standard deviation of Gaussian kernel (Å) | 0.55 | 0.56 | 0.53 | 0.53 | 0.47 | 0.59 | 0.54 | 0.56 |
| Accuracy of atom identification (%) | 96.4 | 96.7 | 97.0 | 97.2 | 94.1 | 94.7 | 96.8 | 96.5 |
| RMSD (pm) | 30.2 | 29.0 | 19.2 | 27.7 | 30.9 | 32.7 | 29.4 | 25.5 |

[a] These nanoparticles correspond to those shown in Fig. 1, while the remaining nanoparticles are presented in Supplementary Fig. 12.

**Supplementary Table 2 | Experimental conditions, analysis parameters and tracing results of Ga-PtNi nanoparticles.**

|  | Ga-PtNi pristine | | | Ga-PtNi 4k | | Ga-PtNi 8k | | Ga-PtNi 12k | |
|---|---|---|---|---|---|---|---|---|---|
|  | Ga-PtNi-pristine-p1[a] | Ga-PtNi-pristine-p2 | Ga-PtNi-pristine-p3 | Ga-PtNi-4k-p1[a] | Ga-PtNi-4k-p2 | Ga-PtNi-8k-p1 | Ga-PtNi-8k-p2[a] | Ga-PtNi-12k-p1 | Ga-PtNi-12k-p2[a] |
| **STEM data acquisition** | | | | | | | | | |
| Electron microscope type | TEAM 0.5 microscope at NCEM | FEI Titan Themis Z at INST | FEI Titan Themis Z at INST | TEAM 0.5 microscope at NCEM | TEAM 0.5 microscope at NCEM | TEAM 0.5 microscope at NCEM | TEAM 0.5 microscope at NCEM | TEAM 0.5 microscope at NCEM | TEAM 0.5 microscope at NCEM |
| Acceleration voltage (kV) | 200 | 300 | 300 | 200 | 200 | 200 | 200 | 200 | 200 |
| Convergence semi-angle (mrad) | 25.0 | 25.1 | 25.1 | 25.0 | 25.0 | 25.0 | 25.0 | 25.0 | 25.0 |
| Detector inner angle (mrad) | 43 | 41 | 41 | 43 | 43 | 43 | 43 | 43 | 43 |
| Detector outer angle (mrad) | 216 | 200 | 200 | 216 | 216 | 216 | 216 | 216 | 216 |
| Pixel size (Å) | 0.329 | 0.357 | 0.357 | 0.329 | 0.329 | 0.329 | 0.329 | 0.329 | 0.329 |
| Screen current (pA) | 10 | 10 | 10 | 10 | 10 | 10 | 10 | 10 | 10 |
| Dwell time (µs) | 3 | 3 | 3 | 3 | 3 | 3 | 3 | 3 | 3 |
| # of consecutive images | 4 | 4 | 4 | 4 | 4 | 4 | 4 | 4 | 4 |
| # of projections | 49 | 38 | 39 | 48 | 48 | 48 | 48 | 51 | 51 |
| Tilt angle range (°) | −65.2 +63.7 | −73.8 +69.2 | −73.0 +73.0 | −66.3 +56.7 | −66.9 +57.1 | −64.5 +57.7 | −65.1 +57.0 | −69.9 +62.6 | −69.9 +63.1 |
| Electron dose ($10^5$ $e$/Å$^2$) | 3.4 | 2.2 | 2.3 | 3.3 | 3.3 | 3.3 | 3.3 | 3.5 | 3.5 |
| **3D reconstruction** | | | | | | | | | |
| Algorithm | GENFIRE | GENFIRE | GENFIRE | GENFIRE | GENFIRE | GENFIRE | GENFIRE | GENFIRE | GENFIRE |
| Interpolation method | DFT | DFT | DFT | DFT | DFT | DFT | DFT | DFT | DFT |
| # of iterations | 1000 | 1000 | 1000 | 1000 | 1000 | 1000 | 1000 | 1000 | 1000 |
| Interpolation radius (pixel) | 0.1 | 0.1 | 0.1 | 0.1 | 0.1 | 0.1 | 0.1 | 0.1 | 0.1 |
| Oversampling ratio | 4 | 4 | 4 | 4 | 4 | 4 | 4 | 4 | 4 |
| **Tracing result** | | | | | | | | | |
| # of atoms | | | | | | | | | |
| Ni | 11433 | 6810 | 7962 | 11578 | 7309 | 4267 | 10539 | 4116 | 4717 |
| Pt | 14714 | 9705 | 12022 | 14587 | 10142 | 7735 | 14451 | 7585 | 8768 |
| B factor (Å$^2$) | | | | | | | | | |
| Ni | 7.3 | 8.1 | 7.8 | 6.6 | 7.2 | 7.6 | 6.6 | 7.4 | 7.3 |
| Pt | 9.2 | 9.1 | 9.2 | 8.3 | 9.1 | 8.6 | 8.3 | 7.9 | 6.8 |
| R factor (%) | 9.8 | 9.6 | 10.9 | 9.2 | 11.2 | 9.8 | 9.3 | 11.3 | 11.4 |
| **Composition** | | | | | | | | | |
| Pt atomic percent (at%) | 56 | 59 | 60 | 56 | 58 | 64 | 58 | 65 | 65 |
| **Precision estimation** | | | | | | | | | |
| Standard deviation of Gaussian kernel (Å) | 0.57 | 0.63 | 0.62 | 0.55 | 0.62 | 0.60 | 0.61 | 0.61 | 0.51 |
| Accuracy of atom identification (%) | 91.7 | 90.1 | 87.9 | 94.5 | 93.3 | 93.2 | 94.3 | 93.9 | 92.7 |
| RMSD (pm) | 38.6 | 39.2 | 41.1 | 34.8 | 33.1 | 30.4 | 34.2 | 32.1 | 31.5 |

[a] These nanoparticles correspond to those shown in Fig. 1, while the remaining nanoparticles are presented in Supplementary Fig. 13.

**Supplementary Table 3 | Chemical composition of PtNi and Ga-PtNi nanoparticles in terms of Pt atomic percentage, obtained from EDS analysis.**

| EDS analysis | PtNi | | Ga-PtNi | | | |
|---|---|---|---|---|---|---|
| | pristine | 12k | pristine | 4k | 8k | 12k |
| Pt atomic percent (at%) | 60 | 64 | 61 | 62 | 62 | 62 |

**Supplementary Table 4 | Sphericity of PtNi nanoparticles obtained from AET results.**

| | PtNi pristine | | PtNi 12k | | | | | |
|---|---|---|---|---|---|---|---|---|
| | PtNi-pristine-p1 | PtNi-pristine-p2 | PtNi-12k-p1 | PtNi-12k-p2 | PtNi-12k-p3 | PtNi-12k-p4 | PtNi-12k-p5 | PtNi-12k-p6 |
| Sphericity | 0.87 | 0.85 | 0.91 | 0.92 | 0.87 | 0.89 | 0.91 | 0.87 |

**Supplementary Table 5 | Sphericity of Ga-PtNi nanoparticles obtained from AET results.**

| | Ga-PtNi pristine | | | Ga-PtNi 4k | | Ga-PtNi 8k | | Ga-PtNi 12k | |
|---|---|---|---|---|---|---|---|---|---|
| | Ga-PtNi-pristine-p1 | Ga-PtNi-pristine-p2 | Ga-PtNi-pristine-p3 | Ga-PtNi-4k-p1 | Ga-PtNi-4k-p2 | Ga-PtNi-8k-p1 | Ga-PtNi-8k-p2 | Ga-PtNi-12k-p1 | Ga-PtNi-12k-p2 |
| Sphericity | 0.88 | 0.88 | 0.87 | 0.83 | 0.80 | 0.80 | 0.86 | 0.86 | 0.78 |

**Supplementary Table 6 | Ga concentration of Ga-PtNi nanoparticles in the pristine state measured from multiple EDS images.**

| | Image 1 | Image 2 | Image 3 | Image 4 | Image 5 | Image 6 | Image 7 | Image 8 |
|---|---|---|---|---|---|---|---|---|
| Ga atomic percent (at%) | 1.37 | 1.41 | 1.11 | 1.31 | 1.05 | 1.02 | 1.44 | 1.14 |
| | Image 9 | Image 10 | Image 11 | Image 12 | Image 13 | Image 14 | Image 15 | |
| Ga atomic percent (at%) | 1.66 | 0.88 | 0.80 | 1.22 | 1.64 | 2.34 | 0.82 | |

**Supplementary Table 7 | Calculated precision of averaged surface strain of Pt atoms for the PtNi nanoparticles.**

| | PtNi pristine | | PtNi 12k | | | | | |
|---|---|---|---|---|---|---|---|---|
| | PtNi-pristine-p1 | PtNi-pristine-p2 | PtNi-12k-p1 | PtNi-12k-p2 | PtNi-12k-p3 | PtNi-12k-p4 | PtNi-12k-p5 | PtNi-12k-p6 |
| Precision (%) | 0.10 | 0.08 | 0.06 | 0.09 | 0.11 | 0.11 | 0.10 | 0.08 |

**Supplementary Table 8 | Calculated precision of averaged surface strain of Pt atoms for the Ga-PtNi nanoparticles.**

| | Ga-PtNi pristine | | | Ga-PtNi 4k | | Ga-PtNi 8k | | Ga-PtNi 12k | |
|---|---|---|---|---|---|---|---|---|---|
| | Ga-PtNi-pristine-p1 | Ga-PtNi-pristine-p2 | Ga-PtNi-pristine-p3 | Ga-PtNi-4k-p1 | Ga-PtNi-4k-p2 | Ga-PtNi-8k-p1 | Ga-PtNi-8k-p2 | Ga-PtNi-12k-p1 | Ga-PtNi-12k-p2 |
| Precision (%) | 0.10 | 0.12 | 0.12 | 0.08 | 0.09 | 0.10 | 0.09 | 0.11 | 0.10 |

**Supplementary References**